\documentclass[11pt,a4paper]{article}

\usepackage[a4paper,margin=27mm,headheight=14pt,headsep=8mm,footskip=13mm]{geometry}
\usepackage{graphicx}
\usepackage{multicol,multirow}
\usepackage{amsmath,amssymb,amsfonts}
\usepackage{mathrsfs}
\usepackage{amsthm}
\usepackage{rotating}
\usepackage[toc,page]{appendix}
\usepackage{ifpdf}
\usepackage[T1]{fontenc}
\usepackage[utf8]{inputenc}
\usepackage[british]{babel}
\usepackage{newtxtext}
\usepackage{newtxmath}
\usepackage{textcomp}
\usepackage{xcolor}
\usepackage{mathtools}
\usepackage{enumitem}
\usepackage{comment}
\usepackage{microtype}
\usepackage{epstopdf}
\usepackage{csquotes}
\usepackage{abstract}
\usepackage{fancyhdr}
\usepackage{setspace}
\usepackage{titlesec}
\usepackage{titling}
\usepackage[backend=biber,style=authoryear-comp,natbib=true,maxcitenames=2,maxbibnames=99,date=year,giveninits=true,terseinits=true,uniquename=init,dashed=false]{biblatex}
\usepackage[colorlinks,allcolors=blue]{hyperref}

\definecolor{inkblue}{HTML}{234D6B}
\definecolor{ruleblue}{HTML}{B8C8D4}
\hypersetup{linkcolor=inkblue,citecolor=inkblue,urlcolor=inkblue}

\pretitle{\begin{center}\vspace*{-1.2em}\color{inkblue}\LARGE\bfseries}
\posttitle{\par\vspace{0.5em}\color{ruleblue}\rule{0.72\textwidth}{0.7pt}\end{center}\vspace{0.8em}}
\preauthor{\begin{center}\large}
\postauthor{\end{center}\vspace{1.2em}}

\titleformat{\section}
  {\normalfont\Large\bfseries\color{inkblue}}{\thesection}{0.75em}{}
\titleformat{\subsection}
  {\normalfont\large\bfseries\color{inkblue}}{\thesubsection}{0.75em}{}
\titleformat{\subsubsection}
  {\normalfont\normalsize\bfseries\color{inkblue}}{\thesubsubsection}{0.75em}{}
\titlespacing*{\section}{0pt}{2.4ex plus 0.8ex minus 0.2ex}{1ex}
\titlespacing*{\subsection}{0pt}{2ex plus 0.6ex minus 0.2ex}{0.7ex}

\fancypagestyle{plain}{%
  \fancyhf{}
  \fancyfoot[C]{\small\color{inkblue}\thepage}

}

\mathtoolsset{showonlyrefs}

\DeclareMathOperator{\Span}{span}

\newcommand{\bx}{\mathbf{x}}
\newcommand{\bk}{\mathbf{k}}
\newcommand{\bX}{\mathbf{X}}

\theoremstyle{plain}
\newtheorem{theorem}{Theorem}[section]
\newtheorem{lemma}[theorem]{Lemma}
\newtheorem{proposition}[theorem]{Proposition}
\newtheorem{corollary}[theorem]{Corollary}

\theoremstyle{definition}

\newtheorem{remark}[theorem]{Remark}

\renewenvironment{proof}[1][Proof]{\par\noindent\textbf{#1.}\ }{\hfill$\square$\par}
\numberwithin{equation}{section}

\begin{document}

\title{Stability of Solitons in the Parametrically Driven and Damped Nonlinear Schr\"odinger Equation in Arbitrary Dimension}
\author{Fernando Carre\~no-Navas\\
\small IMUS and Departamento de An\'alisis Matem\'atico\\
\small Universidad de Sevilla, 41012 Sevilla, Spain\\
\small \texttt{fcarreno@us.es}}
\date{}
\maketitle

\begin{abstract}
We study the spectral stability of soliton solutions of the parametrically driven and damped nonlinear Schr\"odinger (PDDNLS) equation with power-law nonlinearity $|\Psi|^{2\kappa}\Psi$ in $N$ dimensions. We identify two families of stationary solutions, $\Psi_\pm$, whose existence is determined by two conditions: one involving the nonlinearity exponent~$\kappa$ and the spatial dimension~$N$, and another involving the forcing amplitude~$r$ and the damping coefficient~$\rho$. By linearising around these solutions and introducing a convenient parametrisation through~$\varepsilon$, we reduce the stability problem to a dissipationless eigenvalue problem for two Schr\"odinger-type operators, $L_0$ and $L_1$. We prove that $\Psi_-$ is always unstable, while for $\Psi_+$, we carry out a complete spectral analysis by decomposing the problem into angular momentum sectors $\mathcal{S}_l$. We show that modes with $l \geq 2$ generate no instabilities and that no complex quadruplets arise on~$\mathcal{S}_1$. We also derive an analytical instability criterion on $\mathcal{S}_l$ by combining a Vakhitov--Kolokolov-type argument with a local analysis that determines the direction of the eigenvalue splitting near the origin. In addition, we uncover new phenomena at both small and large values of $\kappa$ that were not investigated in earlier one-dimensional studies. Numerical stability diagrams for $N = 1, 2, 3$ are presented.
\end{abstract}

\section{Introduction}\label{sec:introduction}

The parametrically driven and damped nonlinear Schr\"odinger (PDDNLS) equation describes a broad range of physical phenomena. For example, it captures the dynamics of small-amplitude breathers in easy-plane ferromagnets and long Josephson junctions subject to parametric pumping and dissipation \citep{barashenkov:1991,bondila:1995}. It also accounts for the repulsive interaction of double solitons observed in an oscillating water channel \citep{wang:1997}, excited surface-wave solitons \citep{wang:1996,wang:1998}, and non-propagating hydrodynamic solitons under confinement \citep{gordillo:2014}. Time-dependent parametric pumping can create or annihilate solitons \citep{wang:2001}, support soliton bound states \citep{bogdan:2022}, and stabilise damped solitary waves \citep{miles:1984,barashenkov:1991,barashenkov:2001}.

The inclusion of the dissipative term $-i\rho\Psi$ causes NLS solitons to decay. To counteract this effect, we introduce a parametric driving term of the form $r\,e^{2it}\Psi^\star$, which can stabilise stationary solutions of the PDDNLS equation
\begin{equation}\label{NLS}
    i\Psi_t + \Delta\Psi + 2|\Psi|^{2\kappa}\Psi = r\,e^{2it}\Psi^\star - i\rho\Psi,
\end{equation}
where $r > 0$ is the driving amplitude, $\rho \geq 0$ is the dissipation coefficient, $\kappa > 0$ is the nonlinearity exponent, and $\Delta$ denotes the Laplacian operator on~$\mathbb{R}^N$. In physical settings, higher spatial dimensions encompass optical patterns and Bose--Einstein condensates in multi-dimensional traps ($N \geq 2$), as well as nonlinear Faraday resonance in fluids or magnetic materials. The parameter $r$ quantifies the external pump, $\rho$ measures dissipation, and $\kappa$ controls the strength of the focusing nonlinearity; keeping $N$ arbitrary allows one to treat quasi-one-dimensional waveguides, planar patterns, and genuinely three-dimensional localised states within a single framework.

The consideration of $N>1$ changes the problem substantially. Even for the undamped NLS, \citet{berestycki:1983} and \citet{kwong:1989} established the existence and uniqueness of positive radial ground states in $\mathbb{R}^N$, while \citet{weinstein:1985} showed that their orbital stability is governed by the critical threshold $\kappa = 2/N$; see also \citep{sulem:1999}. For the PDDNLS equation, the available rigorous results are much scarcer and remain largely concentrated in one dimension. In dimensions $N \geq 2$, one must additionally cope with the absence of closed-form stationary profiles, the contribution of higher angular momentum sectors, and more intricate spectral interactions, all of which make the stability problem considerably more delicate.

	For $\kappa = 1$ and $N = 1$, \citet{bogdan:1985} found two exact stationary solutions. \citet{barashenkov:1991} showed that one of them is unstable, whereas the other is stable in a region of the $(\rho,r)$--parameter plane. \citet{alexeeva:1999} then identified the collision of two internal modes that produces a complex quadruplet, a mechanism known as \emph{oscillatory instability}. This should be contrasted with exponential instability, where perturbations grow monotonically without oscillating. In \citet{carreno:2024}, we extended the one-dimensional analysis to arbitrary $\kappa > 0$, proving the existence of two stationary solutions $\Psi_\pm$ for all~$\kappa$ and showing that $\Psi_-$ is always unstable, whereas $\Psi_+$ can be stabilised by the combined action of damping and parametric forcing, even for $\kappa \geq 2$, where the unforced NLS soliton is unstable to amplitude blow-up \citep{sulem:1999}. In that regime, stability can be recovered through the resonance of two internal modes on the real axis, a phenomenon referred to as \emph{oscillatory stability}. Stabilisation of NLS solitons has also been achieved by introducing external potentials with supersymmetry and parity-time symmetry \citep{cooper:2017} and through the study of Lie symmetries and conservation laws in related integrable and non-integrable models \citep{MR4986885,MR5032404,MR4909067}.

	Despite these advances, the analysis of the PDDNLS has remained essentially one-dimensional and has covered only a limited range of nonlinearities, typically $\kappa \in [0.5, 3.5]$ \citep{carreno:2024}. Conceptually, a similar problem was addressed by \citet{barashenkov:2002}, who solved it for the specific case of a cubic nonlinearity ($\kappa = 1$, where the standard NLS equation is integrable) using a variational approach and numerical methods to analyse the stability of the ground state soliton. Their work, together with related studies on the interaction of two-dimensional localized structures \citep{clerc:2010}, provided important insights into the dynamics of parametrically driven systems. In our case, the study is substantially more general: it allows for an arbitrary nonlinearity $\kappa > 0$ and arbitrary spatial dimension $N$, thereby including the $\kappa=1$ case as a particular corollary. Moreover, we discover analytically that any possible instability in the system is governed by radial functions, and we obtain explicit analytical results for the instability at low values of the nonlinearity as well as an upper bound for stability. This sector-by-sector stability analysis clarifies the mechanism behind the stability boundaries and makes explicit which part of the spectrum is genuinely new in higher dimensions. The regimes of very small and very large $\kappa$, where qualitatively different behaviour may arise, have therefore remained unexplored. In this paper we study analytically and numerically the spectral stability of stationary solutions of the PDDNLS equation~\eqref{NLS} in arbitrary spatial dimension and for general values of~$\kappa$.

The paper is organised as follows. In Section~\ref{sec:stationary} we derive the two families of stationary solutions of Eq.~\eqref{NLS} in $N$ dimensions and summarise their main properties. Section~\ref{sec:stability} develops the linear stability framework: we linearise around $\Psi_\pm$, introduce the parametrisation in terms of~$\varepsilon$, and derive the dissipationless eigenvalue problem associated with the Schr\"odinger-type operators $L_0$ and $L_1$ defined in~\eqref{L0}--\eqref{L1}. We also establish the relevant symmetry properties, describe the continuous spectrum, and state the basic stability criteria. In Section~\ref{sec:angular_decomposition} we carry out the angular momentum decomposition and the core spectral analysis. There we prove stability for $l \geq 2$ (Theorem~\ref{thm:stability_l_geq_2}), rule out complex quadruplets on $\mathcal{S}_1$ (Proposition~\ref{prop:l1_structure}), and determine the direction of the eigenvalue bifurcations from $\Lambda = 0$ on both $\mathcal{S}_1$ and $\mathcal{S}_0$ (Theorems~\ref{thm:no_splitting},~\ref{thm:direction_splitting_S1},~\ref{thm:bifurcation_L0_eigenvalue},~\ref{thm:degenerate_S1}, and~\ref{thm:bifurcation_S0}; see also Lemma~\ref{condicion_no_degeneracion}), which yields explicit instability intervals (Proposition~\ref{prop:instability_intervals}). Section~\ref{sec:numerics} describes the numerical methods and presents stability diagrams for $N = 1, 2, 3$ and several values of~$\kappa$. It also provides numerical evidence that the $\mathcal{S}_0$ bifurcation occurs before the $\mathcal{S}_1$ bifurcation, implying that the radial sector determines the full stability picture, and discusses the disappearance of the stability region for large~$\kappa$. Section~\ref{sec:summary} closes with a summary of the main results.

\section{Stationary solutions of the PDDNLS in \texorpdfstring{$N$}{N} dimensions}\label{sec:stationary}

We look for stationary solutions of Eq.~\eqref{NLS} whose temporal frequency is locked to half the frequency of the parametric force. 

 The ansatz
\begin{equation}\label{eq:ansatz}
	\Psi(\mathbf{x},t) = \psi(\mathbf{x})\,e^{-i\Theta/2}\,e^{it},
\end{equation}
with $\psi(\bx) > 0$ a real-valued profile and $\Theta$ a constant phase, reduces Eq.~\eqref{NLS} to a stationary problem provided that
\begin{equation}
	\sin\Theta = \frac{\rho}{r}.
\end{equation}
Hence, there are two admissible phase branches:
\begin{equation}\label{eq:phases}
	\Theta_+ = \arcsin\!\left(\frac{\rho}{r}\right), \qquad \Theta_- = \pi - \arcsin\!\left(\frac{\rho}{r}\right),
\end{equation}
defined for $r \geq \rho$. In either case, the profile $\psi$ satisfies the stationary equation
\begin{equation}\label{stationary_NLS}
	-\Delta\psi + \omega_\pm\,\psi - 2\psi^{2\kappa+1} = 0, \qquad \omega_\pm = 1 + r\cos\Theta_\pm = 1 \pm \sqrt{r^2 - \rho^2} > 0.
\end{equation}
 
Since the condition $\omega_- > 0$ is equivalent to $\sqrt{r^2 - \rho^2} < 1$, the branch $\psi_-$ exists only for $\rho < r < \sqrt{1 + \rho^2}$, whereas $\psi_+$ exists for all $r \geq \rho$.

Equation~\eqref{stationary_NLS} has the same form as the stationary equation for the unforced NLS. Consequently, many properties of $\psi_\pm$ follow directly from the classical theory, which we summarise below:

\begin{proposition}\label{prop:properties_psi}
	The profiles $\psi_\pm$ satisfy the following properties:
	\begin{enumerate}[label=(\roman*)]
		\item For $N > 2$, a positive radial ground state exists provided $\kappa < 2/(N-2)$. For $N \leq 2$, no restriction on $\kappa$ is needed \citep{berestycki:1983,kwong:1989}.
		\item The unperturbed NLS ground state is orbitally stable if and only if $\kappa < 2/N$ \citep{weinstein:1985}.
		\item Each profile $\psi_\pm$ belongs to $H^1(\mathbb{R}^N)\cap C^2(\mathbb{R}^N)$ and is radially symmetric and strictly positive \citep{berestycki:1983,gidas:1979,kwong:1989}.
		\item $\psi_\pm(|\mathbf{x}|) \sim c\,|\mathbf{x}|^{(1-N)/2}\,e^{-\sqrt{\omega_\pm}\,|\mathbf{x}|},  |\mathbf{x}| \to \infty,$ for a constant $c > 0$ \citep{kwong:1989,sulem:1999}.
	\end{enumerate}

	Moreover, the functions $\psi_\pm$ satisfy the \textit{Pohozaev identities}, which will play a central role in the analysis developed later. Every $H^2$-solution of~\eqref{stationary_NLS} satisfies
	\begin{align}
		\int_{\mathbb{R}^N} |\nabla\psi|^2\,d\bx & = \frac{\kappa N}{\kappa+1}\int_{\mathbb{R}^N}|\psi|^{2\kappa+2}\,d\bx, \label{poho1}                  \\
		\omega_\pm\int_{\mathbb{R}^N}|\psi|^2\,d\bx        & = \left(2 - \frac{\kappa N}{\kappa+1}\right)\int_{\mathbb{R}^N}|\psi|^{2\kappa+2}\,d\bx. \label{poho2}
	\end{align}
\end{proposition}

It is worth emphasising that the existence of stationary solutions of the PDDNLS is determined by two independent sets of conditions: one involving $\kappa$ and $N$, and another involving $r$ and $\rho$. The former comes from the existence theory for ground states of the unforced NLS, whereas the latter reflects the frequency-locking condition together with the phase relation above.

The two stationary solutions of the PDDNLS equation~\eqref{NLS} are therefore
\begin{equation}\label{eq:psi_pm}
	\Psi_\pm(\mathbf{x},t) = \psi_\pm(\mathbf{x})\,e^{it - i\Theta_\pm/2}.
\end{equation}

For the specific case of dimension $N = 1$, the unique positive solution is given explicitly by
\begin{equation}\label{eq:soliton_1d}
	\psi_\pm(x) = B_\pm\,\operatorname{sech}^{1/\kappa}\!\left(\kappa\sqrt{\omega_\pm}\,x\right), \qquad B_\pm = \left(\frac{\omega_\pm(\kappa+1)}{2}\right)^{1/(2\kappa)}.
\end{equation}
For $\kappa = 1$, this formula agrees with the ones obtained by \citet{barashenkov:1991}. For $\kappa > 0$, these solutions were recently derived by \citet{carreno:2024}. These structural properties provide the starting point for the linear stability analysis in the next section.

\begin{remark}\label{rem:N_greater_1}
For $N > 1$, exact closed-form expressions for the radial ground state profile $\psi_\pm$ are not available in terms of standard elementary functions. In these cases, $\psi_\pm$ is obtained numerically via a sparse matrix approach based on the finite difference method, as described in Section~\ref{sec:numerics}.
\end{remark}

\section{Linear stability analysis}\label{sec:stability}

We linearise Eq.~\eqref{NLS} around its stationary solution by substituting $\Psi(\bx,t) = [\psi_\pm(\bx) + u(\bx,t)]\,e^{it - i\Theta_\pm/2}$ into Eq.~\eqref{NLS}, where $\psi_\pm$ is a ground state of~\eqref{stationary_NLS} and the perturbation $u(\bx,t)$ is small. Decomposing into real and imaginary parts via $u(\bx,t) = e^{-\rho t}(f(\bx,t) + i\,g(\bx,t))$, the linearised system becomes
\begin{align}
	f_t - \rho f  & = \left(-\Delta + 1 - r\cos\Theta_\pm - 2\psi_\pm^{2\kappa}\right)g, \label{eq:lin_f}            \\
	-g_t - \rho g & = \left(-\Delta + 1 + r\cos\Theta_\pm - 2(2\kappa+1)\psi_\pm^{2\kappa}\right)f. \label{eq:lin_g}
\end{align}
With this definition of $u(\bx,t)$, both equations are influenced by the damping in the same way.

A direct stability analysis requires the variation of $\kappa$, $r$, and $\rho$. Following \citet{barashenkov:1991}, we introduce the rescaled variables $\bX = \sqrt{\omega_\pm}\,\bx$, $T = \omega_\pm\,t$, the parameter
\begin{equation}\label{eq:def_eps}
	\varepsilon = 2\!\left(1 - \frac{1}{\omega_\pm}\right), \qquad \widetilde{\rho} = \frac{\rho}{\omega_\pm},
\end{equation}
and denote by $\phi(\bX)$ the rescaled profile satisfying
\begin{equation}\label{eq:psi_rescaled}
	-\Delta\phi + \phi - \phi^{2\kappa+1} = 0,
\end{equation}
so that $\phi(\bX) = (2/\omega_\pm)^{1/(2\kappa)}\,\psi_\pm\!\left(\frac{\bX}{\sqrt{\omega_\pm}}\right).$

Note that $\phi$ is independent of $\varepsilon$.
The parameter ranges over $\varepsilon \in (-\infty, 0)$ for $\Psi_-$ and $\varepsilon \in [0,2)$ for $\Psi_+$. When $\varepsilon = 0$, we have $r = \rho$ and the solution reduces to the soliton of the unforced NLS equation with a specific phase.

In these rescaled variables, the linearised system~\eqref{eq:lin_f}--\eqref{eq:lin_g} takes the form
\begin{align}
	f_T - \widetilde{\rho}\,f  & = L_0\,g, \label{eq:lin_rescaled_f} \\
	-g_T - \widetilde{\rho}\,g & = L_1\,f, \label{eq:lin_rescaled_g}
\end{align}
where the self-adjoint operators $L_0$ and $L_1$ are defined as
\begin{align}
	L_0 & = -\Delta + 1 - \varepsilon - \phi^{2\kappa}, \label{L0} \\
	L_1 & = -\Delta + 1 - (2\kappa+1)\phi^{2\kappa}. \label{L1}
\end{align}
For $N = 1$ these operators agree with those studied by \citet{carreno:2024}.

\subsection{Eigenvalue problem for stability}

We seek solutions of~\eqref{eq:lin_rescaled_f}--\eqref{eq:lin_rescaled_g} of the form
\begin{align}
	f(\bX,T) & = e^{\nu T}\operatorname{Re}\!\left[e^{i\Omega T}f_c(\bX)\right], \label{eq:ansatz_f}             \\
	g(\bX,T) & = e^{\nu T}\operatorname{Re}\!\left[e^{i\Omega T}\widetilde{g}_c(\bX)\right], \label{eq:ansatz_g}
\end{align}
with $\nu, \Omega \in \mathbb{R}$ and $f_c, \widetilde{g}_c$ complex-valued functions not identically zero. The soliton is stable whenever $\nu \leq \widetilde{\rho}$. Setting $\lambda = \nu + i\Omega$, we obtain
\begin{align}
	L_0\,\widetilde{g}_c & = (\lambda - \widetilde{\rho})\,f_c, \label{eq:ev_dissipative_1}              \\
	L_1\,f_c             & = -(\lambda + \widetilde{\rho})\,\widetilde{g}_c. \label{eq:ev_dissipative_2}
\end{align}
When $\lambda = \nu = \widetilde{\rho}$, the soliton is marginally stable. For $\lambda \neq \widetilde{\rho}$, defining
\begin{align}
	\Lambda^2 & = \lambda^2 - \widetilde{\rho}^2, \label{eq:defLambda}                                   \\
	{g}_c     & = \sqrt{\frac{\lambda + \widetilde{\rho}}{\lambda - \widetilde{\rho}}}\,\widetilde{g}_c,
\end{align}
the system reduces to the \emph{dissipationless eigenvalue problem}
\begin{align}
	L_0\,{g}_c & = \Lambda\,f_c, \label{eq:9kappa}     \\
	L_1\,f_c   & = -\Lambda\,{g}_c, \label{eq:9kappaa}
\end{align}
where $\Lambda$ can be complex.

Setting $\Lambda = \Lambda_r + i\Lambda_i$ in Eq.~\eqref{eq:defLambda}, it follows that
\begin{equation}\label{eq:nu_formula}
	\nu^2 = \frac{\Lambda_r^2 - \Lambda_i^2 + \widetilde{\rho}^2}{2} + \frac{\sqrt{(\Lambda_r^2 - \Lambda_i^2 + \widetilde{\rho}^2)^2 + 4\Lambda_r^2\Lambda_i^2}}{2}.
\end{equation}
For the soliton to be stable ($\nu \leq \widetilde{\rho}$), $\widetilde{\rho}$ and $\Lambda$ must satisfy
\begin{equation}
	\widetilde{\rho}^2(\Lambda_i^2 - \Lambda_r^2) \geq \Lambda_r^2\Lambda_i^2.
\end{equation}

A necessary condition for stability is therefore
\begin{equation}\label{neccesary_condition_stability}
	|\Lambda_i| > |\Lambda_r|.
\end{equation}
In particular, if $\Lambda = i\Lambda_i$ is purely imaginary, it never produces instability. However, if $\Lambda = \Lambda_r \neq 0$ is real, the soliton is always unstable.

Assume that for a given $\varepsilon$, a complex eigenvalue $\Lambda = \Lambda_r + i\Lambda_i$ appears with $\Lambda_i > \Lambda_r > 0$. The threshold of instability occurs at $\nu = \widetilde{\rho}$, giving
\begin{equation}\label{eq:rho_eps}
	\rho(\varepsilon) = \frac{2\nu}{2-\varepsilon}, \qquad \nu = \frac{\Lambda_i\Lambda_r}{\sqrt{\Lambda_i^2 - \Lambda_r^2}},
\end{equation}
where the relation $\widetilde{\rho} = \rho/\omega_\pm$ has been used. From the definition of $\varepsilon$, the amplitude of the parametric force can be expressed as
\begin{equation}\label{eq:stability_curve}
	r(\varepsilon) = \sqrt{\rho^2(\varepsilon) + \left(\frac{\varepsilon}{2-\varepsilon}\right)^2}.
\end{equation}
The stability curve is formed by all points $(\rho(\varepsilon), r(\varepsilon))$, where $\varepsilon_c \leq \varepsilon < 2$. Here $\varepsilon_c$ denotes the critical value of $\varepsilon$ for which a complex eigenvalue with $|\Lambda_i| > |\Lambda_r|$ first arises. This analysis holds for arbitrary $\kappa$ and~$N$.

\subsection{Instability of \texorpdfstring{$\Psi_-$}{Psi-}}\label{sec:psi_minus_instability}

We now show that $\Psi_-$ is always unstable, regardless of $\kappa > 0$ and $N \geq 1$, by proving that the eigenvalue problem~\eqref{eq:9kappa}--\eqref{eq:9kappaa} with $\varepsilon  < 0$ possesses at least one real positive eigenvalue. For $\kappa = 1$ and $N = 1$ this was proved by \citet{barashenkov:1990}, and for arbitrary $\kappa$ and $N=1$ it was shown by \citet{carreno:2024}.

\begin{proposition}\label{prop:L0_positive}
	For $\varepsilon < 0$, the operator $L_0$ is strictly positive definite.
\end{proposition}

\begin{proof}
	From Eq.~\eqref{eq:psi_rescaled}, $L_0\phi = -\varepsilon\phi$. Since $\phi > 0$ is nodeless, $-\varepsilon > 0$ is the lowest eigenvalue of $L_0$ (see \citep{morse:1953}). Therefore $L_0 > 0$.
\end{proof}

\begin{theorem}\label{thm:psi_minus}
	The solution $\Psi_-$ of Eq.~\eqref{NLS} is spectrally unstable for all $\kappa > 0$ and $N \geq 1$.
\end{theorem}

\begin{proof}
	Since $L_0$ is positive definite and hence invertible by Proposition~\ref{prop:L0_positive}, the system~\eqref{eq:9kappa}--\eqref{eq:9kappaa} can be rewritten as
	\begin{equation}\label{eq:L1_pencil}
		L_1 f_c = -\Lambda^2 L_0^{-1} f_c,
	\end{equation}
	giving
	\begin{equation}\label{eq:Lambda_squared}
		\Lambda^2 = -\frac{\langle f_c, L_1 f_c\rangle}{\langle f_c, L_0^{-1} f_c\rangle}.
	\end{equation}
	Define the functional $\Sigma({\xi}) = -\langle\xi, L_1\xi\rangle / \langle\xi, L_0^{-1}\xi\rangle$. Writing $f_c = L_0^{1/2}\eta$, problem~\eqref{eq:L1_pencil} becomes
	\[
		A\eta = -\Lambda^2\eta, \qquad A := L_0^{1/2}L_1L_0^{1/2}.
	\]
	Moreover,
	\[
		\Sigma(f_c) = -\frac{\langle \eta, A\eta\rangle}{\|\eta\|_{L^2}^2}.
	\]
	Using $L_1 = L_0 + \varepsilon - 2\kappa\phi^{2\kappa}$, we can write
	\[
		A = L_0^2 + \varepsilon L_0 - 2\kappa L_0^{1/2}\phi^{2\kappa}L_0^{1/2}.
	\]
	Since $\phi^{2\kappa} \in L^\infty(\mathbb{R}^N)$, for any $u \in D(L_0)$ (domain of $L_0$) we have,
	\[
		\langle u, Au\rangle \ge \|L_0u\|_{L^2}^2 - C\langle u, L_0u\rangle,
		\qquad C := 2\kappa\|\phi\|_{L^\infty}^{2\kappa} - \varepsilon.
	\]
	Since $L_0 \ge 0$, the spectral theorem gives $L_0^2 - CL_0 \ge -\tfrac{C^2}{4}I$, and therefore $A$ is bounded from below.

	Next, let $A_0 := L_0^2 + \varepsilon L_0$. Because $\phi$ decays exponentially, multiplication by $\phi^{2\kappa}$ is compact from $H^1(\mathbb{R}^N)$ to $H^{-1}(\mathbb{R}^N)$, so the form perturbation $L_0^{1/2}\phi^{2\kappa}L_0^{1/2}$ is compact relative to $A_0$. By Weyl's theorem, $\sigma_{\mathrm{ess}}(A) = \sigma_{\mathrm{ess}}(A_0)$. Since $L_0$ is a Schr\"odinger operator with exponential decaying potential and $\varepsilon < 0$, we have $\sigma_{\mathrm{ess}}(L_0) = [1-\varepsilon,+\infty)$, and therefore
	\[
		\sigma_{\mathrm{ess}}(A_0) = \{t^2+\varepsilon t : t \in [1-\varepsilon,+\infty)\} = [1-\varepsilon,+\infty) \subset (0,+\infty).
	\]

	On the other hand, $L_1$ has a negative eigenvalue with nodeless eigenfunction $\phi_0$ (Lemma~\ref{lem:L1_spectral} below). Setting $\eta_0 := L_0^{-1/2}\phi_0$, we obtain
	\[
		\langle \eta_0, A\eta_0\rangle = \langle \phi_0, L_1\phi_0\rangle < 0.
	\]
	Hence $\inf\sigma(A) < 0$. Since the essential spectrum of $A$ is positive, $\inf\sigma(A)$ is an isolated negative eigenvalue; write it as $-\Lambda_0^2$ with $\Lambda_0 > 0$. Then
	\[
		\Lambda_0^2 = \sup_{\xi \neq 0}\Sigma(\xi),
	\]
	and the supremum is attained by an eigenfunction of $A$. Equivalently, there exists $f_0 \neq 0$ such that
	\[
		(L_1 + \Lambda_0^2 L_0^{-1})f_0 = 0.
	\]
	Thus $\Lambda_0 > 0$ is a real eigenvalue of~\eqref{eq:L1_pencil}, and $\Psi_-$ is always unstable.
\end{proof}
\medskip
From now on we restrict attention to the $\Psi_+$ branch.


\subsection{Symmetries and Continuous Spectra}

\begin{proposition}[Spectral symmetries]\label{prop:spectral_symmetries}
	The spectrum of the eigenvalue problem~\eqref{eq:9kappa}--\eqref{eq:9kappaa} is symmetric with respect to both the real and imaginary axes. More precisely:
	\begin{enumerate}
		\item[\textup{(i)}] If $\Lambda$ is an eigenvalue with eigenfunction $(f_c, g_c)$, then $\Lambda^\star$ is an eigenvalue with eigenfunction $(f_c^\star, g_c^\star)$.
		\item[\textup{(ii)}] If $\Lambda$ is an eigenvalue with eigenfunction $(f_c, g_c)$, then $-\Lambda$ is an eigenvalue with eigenfunction $(f_c, -g_c)$.
	\end{enumerate}
	Consequently, every complex eigenvalue $\Lambda = \Lambda_r + i\Lambda_i$ with $\Lambda_r\Lambda_i \neq 0$ belongs to a quartet $\{\Lambda, -\Lambda, \Lambda^\star, -\Lambda^\star\}$.
\end{proposition}

\begin{proof}
	
	Property~(i) follows immediately from the fact that $L_0$ and $L_1$ are real operators, while~(ii) follows from substituting $(g_c, \Lambda) \to (-g_c, -\Lambda)$ into~\eqref{eq:9kappa}--\eqref{eq:9kappaa}.
\end{proof}

The stability of the soliton $\Psi_+$ is governed not only by the discrete eigenvalues of~\eqref{eq:9kappa}--\eqref{eq:9kappaa}, but also by the continuous spectrum, which arises from the behaviour of the linearised operators at spatial infinity.

\begin{proposition}[Continuous spectrum]\label{prop:continuous_spectrum}
	The continuous spectrum of the eigenvalue problem~\eqref{eq:9kappa}--\eqref{eq:9kappaa} is determined by the dispersion relation
	\begin{equation}\label{eq:dispersion}
		\Lambda_c(\mathbf{k}) = \pm i\sqrt{(1 + |\mathbf{k}|^2)(1 + |\mathbf{k}|^2 - \varepsilon)}, \qquad \mathbf{k} \in \mathbb{R}^N,
	\end{equation}
	and is independent of~$\kappa$. In particular:
	\begin{enumerate}
		\item[\textup{(i)}] If $0 \leq \varepsilon \leq 1$, then $\Lambda_c(\mathbf{k})$ is purely imaginary for all $\mathbf{k} \in \mathbb{R}^N$ and the continuous spectrum lies on the imaginary axis:
		      \begin{equation}\label{eq:cont_spec_stable}
			      \sigma_c = \bigl(-i\infty,\, -i\sqrt{1-\varepsilon}\,\bigr] \cup \bigl[\,i\sqrt{1-\varepsilon},\, +i\infty\bigr).
		      \end{equation}
		\item[\textup{(ii)}] If $1 < \varepsilon < 2$, then there exist wavevectors $\mathbf{k}$ with $|\mathbf{k}|^2 < \varepsilon - 1$ for which $\Lambda_c(\mathbf{k}) \in \mathbb{R}\setminus\{0\}$, so the continuous spectrum contains a real segment and the soliton $\Psi_+$ is unstable.
	\end{enumerate}
	Therefore, stability of $\Psi_+$ requires $0 \leq \varepsilon \leq 1$, equivalently $\rho \leq r \leq \sqrt{1 + \rho^2}$.
\end{proposition}

\begin{proof}
	Since $\phi(\bx) \to 0$ exponentially as $|\bx| \to \infty$ (property~(iv) of Proposition~\ref{prop:properties_psi}), the operators~\eqref{L0}--\eqref{L1} reduce to their asymptotic forms, due to Weyl's theorem on the invariance of the essential spectrum under compact perturbations \citep{teschl:2009}:
	\begin{equation*}
		L_0^\infty = -\Delta + 1 - \varepsilon, \qquad L_1^\infty = -\Delta + 1.
	\end{equation*}
	The continuous spectrum is obtained by the application of the Fourier transform to the system
	\begin{equation}
		\begin{pmatrix}
			0           & L_0^\infty \\
			-L_1^\infty & 0
		\end{pmatrix},
	\end{equation}
	with $L_0$ and $L_1$ replaced by $L_0^\infty$ and $L_1^\infty$, respectively. This gives
	\begin{equation}\label{compatibility_condition}
		(|\bk|^2 + 1 - \varepsilon)\,\hat{g} = \Lambda\,\hat{f}, \qquad (|\bk|^2 + 1)\,\hat{f} = -\Lambda\,\hat{g},
	\end{equation}
	where $\hat{f}$ and $\hat{g}$ are the Fourier transforms of $f$ and $g$. Equation \eqref{compatibility_condition} yields the dispersion relation~\eqref{eq:dispersion}.

	\smallskip\noindent\textup{(i)} For $0 \leq \varepsilon \leq 1$, both factors $(1 + |\mathbf{k}|^2) \geq 1$ and $(1 + |\mathbf{k}|^2 - \varepsilon) \geq 1 - \varepsilon \geq 0$ are non-negative for all $\mathbf{k}$, so $\Lambda_c(\mathbf{k})^2 \leq 0$ and $\Lambda_c$ is purely imaginary. The minimum modulus $|\Lambda_c| = \sqrt{1-\varepsilon}$ is attained at $\mathbf{k} = 0$, while $|\Lambda_c| \to \infty$ as $|\mathbf{k}| \to \infty$, giving~\eqref{eq:cont_spec_stable}.

	\smallskip\noindent\textup{(ii)} For $1 < \varepsilon < 2$, the factor $(1 + |\mathbf{k}|^2 - \varepsilon)$ becomes negative when $|\mathbf{k}|^2 < \varepsilon - 1$, making $\Lambda_c(\mathbf{k})^2 > 0$. For these values, $\Lambda_c(\mathbf{k}) \in \mathbb{R} \setminus \{0\}$. Since a real eigenvalue $\Lambda_r \neq 0$ always produces $\nu > \widetilde{\rho}$ (cf.~\eqref{neccesary_condition_stability}), the soliton is unstable.

	\smallskip
	Combining~(i) and~(ii): the continuous spectrum contributes no instability if and only if $0 \leq \varepsilon \leq 1$. Using $\varepsilon = 2(1 - 1/\omega_+)$ and $\omega_+ = 1 + \sqrt{r^2 - \rho^2}$, the condition $\varepsilon \leq 1$ is equivalent to $\omega_+ \leq 2$, i.e., $r \leq \sqrt{1 + \rho^2}$, while $\varepsilon \geq 0$ corresponds to $r \geq \rho$.
\end{proof}

Proposition~\ref{prop:continuous_spectrum} implies that the soliton $\Psi_+$ is unstable for $\varepsilon > 1$, and hence for $r > \sqrt{1 + \rho^2}$, due to the presence of a real segment in the continuous spectrum. For $0 \leq \varepsilon \leq 1$, the continuous spectrum lies on the imaginary axis and does not contribute to instability, so the stability of $\Psi_+$ is determined solely by the discrete eigenvalues of~\eqref{eq:9kappa}--\eqref{eq:9kappaa}.

\section{Spectral analysis by angular momentum decomposition}\label{sec:angular_decomposition}

In $N$ spatial dimensions, the radial symmetry of the ground state allows us to decompose the linearised eigenvalue problem~\eqref{eq:9kappa}--\eqref{eq:9kappaa} into angular momentum sectors, providing a complete spectral analysis.
\subsection{Decomposition into angular momentum subspaces}

We decompose the Laplacian in hyperspherical coordinates:
\begin{equation}
	\Delta = \frac{1}{r^{N-1}}\frac{\partial}{\partial r}\left(r^{N-1}\frac{\partial}{\partial r}\right) + \frac{1}{r^2}\Delta_{S^{N-1}},
\end{equation}
where $\Delta_{S^{N-1}}$ is the Laplace--Beltrami operator on $S^{N-1}$ with $r = |\bx|$. The hyperspherical harmonics $Y_l^m$ satisfy $\Delta_{S^{N-1}} Y_l^m = -l(l+N-2)Y_l^m$ \citep{avery:2000} for $N\geq 2$. We will consider the convention that for $N=1$, the subspace corresponding to $l=0$ denotes even functions and the subspace for $l=1$ denotes odd functions.

For each integer $l \geq 0$, define the angular momentum subspace
\begin{equation}\label{def:S_l}
	\mathcal{S}_l := \bigg\{u \in H^1(\mathbb{R}^N) : u(\bx) = \sum_m u_{l,m}(r)\,Y_l^m(\hat{\bx})\bigg\},
\end{equation}
where $m$ ranges over the $d_l$ independent harmonics with
\begin{equation}
	d_l = \frac{(2l+N-2)(l+N-3)!}{l!\,(N-2)!}, \qquad l \geq 1, \qquad d_0 = 1
\end{equation}
(for $N = 3$, $d_l = 2l+1$) and $\hat{\bx} = \bx/r$. Since the potentials in $L_0$ and $L_1$ are radially symmetric, both operators commute with $\Delta_{S^{N-1}}$ and hence leave each $\mathcal{S}_l$ invariant. We denote the restrictions by $L_0^l := L_0\big|_{\mathcal{S}_l}$ and $L_1^l := L_1\big|_{\mathcal{S}_l}$.

Expanding $g(\bx) = \sum_l g_l(r)\,Y_l(\hat{\bx})$ and $f(\bx) = \sum_l f_l(r)\,Y_l(\hat{\bx})$, the system~\eqref{eq:9kappa}--\eqref{eq:9kappaa} decouples into a radial problem for each $l\geq0$:
\begin{align}
	-\frac{1}{r^{N-1}}\frac{\partial}{\partial r}\!\left(r^{N-1}\frac{\partial g_l}{\partial r}\right) + \left[\frac{l(l+N-2)}{r^2} + (1-\varepsilon) - \phi^{2\kappa}\right]g_l & = \Lambda\,f_l, \label{eq:radial_system_1}  \\
	-\frac{1}{r^{N-1}}\frac{\partial}{\partial r}\!\left(r^{N-1}\frac{\partial f_l}{\partial r}\right) + \left[\frac{l(l+N-2)}{r^2} + 1 - (2\kappa+1)\phi^{2\kappa}\right]f_l    & = -\Lambda\,g_l, \label{eq:radial_system_2}
\end{align}
with regularity at the origin: $g_l(r) = O(r^l)$, $f_l(r) = O(r^l)$ as $r \to 0^+$.

We also introduce the shifted operator
\begin{equation}\label{def:tildeL0}
	\tilde{L}_0 := L_0 + \varepsilon = -\Delta + 1 - \phi^{2\kappa},
\end{equation}
and its restrictions $\tilde{L}_0^l := \tilde{L}_0\big|_{\mathcal{S}_l}$.

Throughout the rest of the paper,  we denote by $\lambda_i^{(l)}$ the $i$-th discrete eigenvalue of $\tilde{L}_0^l$, listed in increasing order whenever it exists.

\subsection{Spectral properties of the restricted operators}

We collect the fundamental spectral properties of $L_1$ and $\tilde{L}_0$ that underpin the entire stability analysis.

\begin{lemma}[Spectral structure of $L_1$]\label{lem:L1_spectral}
	Assume $0 < \kappa < 2/(N-2)$ for $N\geq 3$ and no restriction on $\kappa$ when $N \leq 2$. Then, the operator $L_1$ acting on $L^2(\mathbb{R}^N)$ satisfies:
	\begin{enumerate}[label=(\roman*)]
		\item $L_1$ possesses exactly one simple negative eigenvalue, whose eigenfunction is radial and lies in~$\mathcal{S}_0$.
		\item $\ker(L_1) = \operatorname{span}\{\partial_{x_j}\phi : j = 1, \ldots, N\} \subset \mathcal{S}_1$.
		\item The restriction $L_1^l$ is strictly positive for all $l \geq 2$.
		\item If $\kappa < 2/N$, then there exists $C>0$ such that $\langle h, L_1 h\rangle \geq C \langle \nabla h, \nabla \phi \rangle^2$ for all $h \in \{\phi\}^\perp$.
		\item There exists $C>0$ such that $\langle h, L_1 h\rangle \geq C \langle  h, \phi \rangle^2$ for all $h \in \{\phi^{2\kappa+1}\}^\perp$.
	\end{enumerate}
\end{lemma}

\begin{proof}
	\emph{(i):} This is a classical result due to \citet{weinstein:1985}. The function $\phi^{\kappa+1}$ is radial and positive, and one verifies $\langle\phi^{\kappa+1}, L_1\phi^{\kappa+1}\rangle < 0$ using the Pohozaev identities~\eqref{poho1}--\eqref{poho2}. By the min-max characterisation and the fact that $\phi^{\kappa+1}$ has no nodes, this yields exactly one negative eigenvalue, which is simple and whose eigenfunction lies in~$\mathcal{S}_0$.

	\emph{(ii):} Differentiating the stationary equation $-\Delta\phi + \phi - \phi^{2\kappa+1} = 0$ with respect to $x_j$ yields $L_1(\partial_{x_j}\phi) = 0$. Since $\partial_{x_j}\phi = \phi'(r)\,\hat{x}_j$ with $\hat{x}_j = x_j/r$, these functions lie in $\mathcal{S}_1$ and are linearly independent for $j = 1, \ldots, N$. The other inclusion $\ker(L_1) \subset \operatorname{span}\{\partial_{x_j}\phi\}$ can be found in \citet{chang:2007}.

	\emph{(iii):} Since the unique eigenfunction associated with the negative eigenvalue resides in $\mathcal{S}_0$ and the kernel lies in $\mathcal{S}_1$, the restriction to $\mathcal{S}_l$ with $l \geq 2$ inherits neither negative eigenvalues nor a non-trivial kernel, hence $L_1^l > 0$. At $l = 1$, the kernel is non-trivial but no negative eigenvalue enters this subspace, so $L_1^1 \geq 0$.

	\emph{(iv)--(v):} For the proof, see \citet{weinstein:1985}.
\end{proof}

\begin{lemma}[Spectral structure of $\tilde{L}_0$]\label{lem:L0_spectral}
	Under the same hypotheses as Lemma~\ref{lem:L1_spectral}:
	\begin{enumerate}[label=(\roman*)]
		\item $\tilde{L}_0 \geq 0$ on $L^2(\mathbb{R}^N)$, with $\ker(\tilde{L}_0) = \operatorname{span}\{\phi\} \subset \mathcal{S}_0$.
		\item The restriction $\tilde{L}_0^l$ is strictly positive for all $l \geq 1$.
	\end{enumerate}
\end{lemma}

\begin{proof}
	\emph{(i):} From the stationary equation, $\tilde{L}_0\phi =\phi^{2\kappa+1} - \phi^{2\kappa}\phi = 0$. Since $\phi > 0$ is an eigenfunction with eigenvalue~$0$, the Perron--Frobenius principle \citep{reed:1978} guarantees that $0$ is the lowest eigenvalue and it is simple. Therefore $\tilde{L}_0 \geq 0$.

	\emph{(ii):} Since $\ker(\tilde{L}_0) = \operatorname{span}\{\phi\} \subset \mathcal{S}_0$ and $\tilde{L}_0 \geq 0$, the restriction to any $\mathcal{S}_l$ with $l \geq 1$ has strictly positive spectrum.
\end{proof}

Throughout the rest of Section~\ref{sec:angular_decomposition}, the analytical arguments rely on the existence of the positive radial ground state $\phi$, namely $\kappa < 2/(N-2)$ when $N \geq 3$ and no upper restriction when $N \leq 2$. These results characterise the onset of instability through splitting at $\Lambda = 0$ and through the ordering of the angular sectors, but they do not by themselves locate every oscillatory-instability threshold; the remaining boundaries are therefore completed numerically in Section~\ref{sec:numerics}.

\subsection{Absence of instabilities for \texorpdfstring{$l \geq 2$}{l >= 2}}

\begin{theorem}[Stability for higher angular momentum modes]\label{thm:stability_l_geq_2}
	Under the hypotheses of Lemma~\ref{lem:L1_spectral},  for all $\varepsilon \in [0,1)$ and all $l \geq 2$, the eigenvalue problem~\eqref{eq:9kappa}--\eqref{eq:9kappaa} restricted to $\mathcal{S}_l$ has $\Lambda^2 < 0$. That is, the $\mathcal{S}_l$-component of the spectrum produces no instability for $l \geq 2$.
\end{theorem}

\begin{proof}
	\textit{Step~1: Eigenvalues are real or purely imaginary.}
	Let $(\Lambda, g_l, f_l)$ be an eigentriple with $g_l, f_l \in \mathcal{S}_l \setminus \{0\}$. Taking $L^2$-inner products of~\eqref{eq:9kappa} with $g_l$ and of~\eqref{eq:9kappaa} with $f_l$ and using self-adjointness, we obtain
	\begin{equation}\label{eq:hermitian_products}
		\langle g_l, L_0^l g_l\rangle = \Lambda\langle g_l, f_l\rangle, \qquad \langle f_l, L_1^l f_l\rangle = -\Lambda\langle f_l, g_l\rangle.
	\end{equation}
	The left-hand sides are real. Writing $\langle g_l, f_l\rangle = \alpha + i\beta$ and $\Lambda = a + ib$, reality of $\Lambda\langle g_l, f_l\rangle$ gives $a\beta + b\alpha = 0$, and reality of $\Lambda\langle f_l, g_l\rangle = \Lambda(\alpha - i\beta)$ gives $a\beta - b\alpha = 0$. Adding: $a\beta = 0$; subtracting: $b\alpha = 0$.

	By Lemma~\ref{lem:L1_spectral}(iii), $\langle f_l, L_1^l f_l\rangle > 0$, which forces $\Lambda\langle f_l, g_l\rangle \neq 0$, so $(\alpha,\beta) \neq (0,0)$. If both $\alpha \neq 0$ and $\beta \neq 0$, then $a = b = 0$, contradicting $\Lambda \langle f_l, g_l\rangle\neq 0$. Therefore either $\alpha = 0$ ($\Lambda$ purely imaginary) or $\beta = 0$ ($\Lambda$ real). In summary, $\Lambda^2 \in \mathbb{R}\setminus\{0\}$.

	\textit{Step~2: At $\varepsilon = 0$, all eigenvalues are purely imaginary.}
	When $\varepsilon = 0$, $L_0 = \tilde{L}_0$, and by Lemma~\ref{lem:L0_spectral}(ii), $\tilde{L}_0^l > 0$ for $l \geq 1$. Hence $\langle g_l, L_0^l g_l\rangle > 0$, giving $\Lambda^2 < 0$.

	\textit{Step~3: Continuity argument.}
	The eigenvalues depend continuously on $\varepsilon$ \citep[Chapter~IV]{kato:1995}. Since by Step~1 eigenvalues must remain on the real or imaginary axis, a transition from purely imaginary to real requires passing through $\Lambda = 0$, which cannot happen. Therefore the purely imaginary character is preserved for all $\varepsilon \in [0,1)$.
\end{proof}

\begin{remark}\label{rem:l2_loophole}
	The essential spectrum of $L_0^l$ is $[1-\varepsilon, +\infty)$, so the gap between discrete eigenvalues and the essential spectrum persists for $\varepsilon \in [0,1)$. At $\varepsilon = 1$, the essential spectrum touches the origin but does not produce instability by the same argument as in Theorem~\ref{thm:stability_l_geq_2}.
\end{remark}

\subsection{The case \texorpdfstring{$l = 1$}{l = 1}: absence of complex quadruplets}

\begin{proposition}[Structure of eigenvalues on $\mathcal{S}_1$]\label{prop:l1_structure}
	For $l = 1$, the eigenvalues $\Lambda$ of the restricted system~\eqref{eq:9kappa}--\eqref{eq:9kappaa} satisfy $\Lambda^2\in \mathbb{R}$. No complex eigenvalue quadruplets can appear in~$\mathcal{S}_1$.
\end{proposition}

\begin{proof}
	By Lemma~\ref{lem:L1_spectral}(iv), $L_1^1 \geq 0$, so $\langle f_l, L_1^1 f_l\rangle \geq 0$ and the same computation as in Step~1 of Theorem~\ref{thm:stability_l_geq_2} yields $\Lambda^2 \in \mathbb{R}$.
\end{proof}

Unlike the $l \geq 2$ case, the semidefiniteness (rather than strict positivity) of $L_1^1$ means that the real and imaginary axes are connected through the origin. A pair of purely imaginary eigenvalues can coalesce at $\Lambda = 0$ and subsequently split into a pair of real eigenvalues. The conditions under which such transitions occur are characterised in Theorem~\ref{thm:no_splitting} and Proposition~\ref{prop:splitting_criterion}.

\subsection{Splitting analysis at zero}\label{sec:splitting_zero}

We characterise the conditions under which eigenvalues can bifurcate from $\Lambda = 0$.

Define the matrix operator
\begin{equation}\label{def:calL}
	\mathcal{L} := \begin{pmatrix} 0 & L_0 \\ -L_1 & 0 \end{pmatrix},
\end{equation}
so that the eigenvalue problem~\eqref{eq:9kappa}--\eqref{eq:9kappaa} can be written as $\mathcal{L}(f,g)^T = \Lambda(f,g)^T$.

\begin{theorem}[Necessary splitting condition for $\varepsilon \notin \sigma(\tilde{L}_0)$]\label{thm:no_splitting}

	Under the hypotheses of Lemma~\ref{lem:L1_spectral}, suppose  $\varepsilon \notin \sigma(\tilde{L}_0)$, so that $L_0 = \tilde{L}_0 - \varepsilon$ is invertible, and let
	\[
		\mu(\varepsilon) := \langle\partial_{x_1}\phi, L_0^{-1}\partial_{x_1}\phi\rangle.
	\]
	Then:
	\begin{itemize}
		\item $\mu(\varepsilon)$ is a strictly increasing function of $\varepsilon$.
		\item $\mu(\varepsilon)$ diverges at those values of $\varepsilon \in \sigma(\tilde{L}_0)$ for which $\partial_{x_1}\phi \notin \ker(L_0)^\perp$.
		\item If $\mu(\varepsilon) \neq 0$, then the generalised null space of $\mathcal{L}$ has dimension exactly $2N$.
		\item If $\mu(\varepsilon) = 0$, then the generalised null space of $\mathcal{L}$ extends to dimension $\geq4N$.
		\item In particular, if $\varepsilon < \lambda_0^{(1)}$ (when it exists), then $\mu(\varepsilon) > 0$ and no splitting occurs.
	\end{itemize}
\end{theorem}

\begin{proof}
	\textit{Step~1: Structure of $\ker(\mathcal{L})$.}
	Setting $\Lambda = 0$ gives $L_0 g = 0$ and $L_1 f = 0$. Since $L_0$ is invertible, $g = 0$. By Lemma~\ref{lem:L1_spectral}(ii), $f \in \ker(L_1) = \operatorname{span}\{\partial_{x_j}\phi\}_{j=1}^N$. Hence $\dim\ker(\mathcal{L}) = N$.

	\textit{Step~2: Calculation of $\ker(\mathcal{L}^2)$.}
	We seek  $(f,g)^T$ with  $\mathcal{L}(f,g)^T$ $\in \ker(\mathcal{L})$. This requires $f \in \ker(L_1)$ and $L_0 g = \sum_j c_j\partial_{x_j}\phi$. Therefore $\dim\ker(\mathcal{L}^2) = 2N$.

	\textit{Step~3: Obstruction to extending the Jordan chain.}
	For $\ker(\mathcal{L}^3) \supsetneq \ker(\mathcal{L}^2)$, we need $L_1 f_2 = -L_0^{-1}\partial_{x_k}\phi$. By the Fredholm alternative, a solution exists if and only if $M_{jk} := \langle\partial_{x_j}\phi, L_0^{-1}\partial_{x_k}\phi\rangle = 0$ for all~$j$ and $\dim(\ker(\mathcal{L}^3)) = 4N$.

	\textit{Step~4: The matrix $M_{jk}$ is diagonal.}
	Since the subspaces of $\mathcal{S}_1$ corresponding to different values of $j$ are preserved by $L_0$ and are mutually orthogonal, $M_{jk} = 0$ for $j \neq k$ by angular orthogonality. By rotational symmetry, $M_{jj} = \mu$ for all~$j$.

	\textit{Step~5: Proof that $\mu$ is a strictly increasing function.}

	We define $\chi = L_0^{-1}(\partial_{x_1}\phi) \in \mathcal{S}_1$, so that $L_0\chi = \partial_{x_1}\phi$. Thus,
	\begin{equation}
		\mu(\varepsilon) = \langle \partial_{x_1}\phi, \chi \rangle.
	\end{equation}
	Differentiating with respect to $\varepsilon$ gives
	\begin{equation}\label{mu_derivative}
		\mu'(\varepsilon) = \langle \partial_{x_1}\phi, \chi' \rangle.
	\end{equation}
	Differentiating $L_0\chi = \partial_{x_1}\phi$ with respect to $\varepsilon$, we obtain $L_0\chi' + L_0'\chi = 0$. Since $L_0' = -I$, we have $L_0\chi' = \chi$. Substituting into \eqref{mu_derivative} gives
	\begin{equation}
		\mu'(\varepsilon) = \langle \partial_{x_1}\phi, L_0^{-1}\chi \rangle = \langle \partial_{x_1}\phi, L_0^{-2}\partial_{x_1}\phi \rangle > 0,
	\end{equation}
	because $L_0^{-2}$ is a positive operator.

	\textit{Step~6: Proof that $\mu > 0$ for $\varepsilon<\lambda_0^{(1)}$.}
	At this regime, $L_0^1 = \tilde{L}_0^1 - \varepsilon$ is itself a positive operator on $\mathcal{S}_1$ and
	\begin{equation}\label{eq:mu_small_eps}
		\mu(\varepsilon) = \langle L_0\chi, \chi \rangle \geq \bigl(\lambda_0^{(1)} - \varepsilon\bigr)\|\chi\|^2 > 0.
	\end{equation}
\end{proof}

\subsection{Splitting conditions on \texorpdfstring{$\mathcal{S}_0$}{S0} and \texorpdfstring{$\mathcal{S}_1$}{S1}}

When $\varepsilon \in \sigma(\tilde{L}_0)$, the operator $L_0$ is no longer invertible and new phenomena occur.

\begin{proposition}[Necessary condition for splitting]\label{prop:splitting_criterion}
	Suppose $\varepsilon \in \sigma(\tilde{L}_0^l)$ for $l \in \{0,1\}$, so that $\ker(L_0^l) \neq \{0\}$. A necessary condition for splitting at $\Lambda = 0$ is the existence of $\zeta \in \ker(L_0^l)\setminus\{0\}$ satisfying
	\begin{equation}\label{eq:splitting_iff}
		\langle\zeta, \partial_{x_j}\phi\rangle = 0 \qquad \text{for all } j = 1, \ldots, N.
	\end{equation}
\end{proposition}

\begin{proof}
	When $\varepsilon \in \sigma(\tilde{L}_0^l)$, the vector $(\zeta, 0)^T \in \ker(\mathcal{L})$. The Jordan chain can be extended if there exists $f_1$ with $L_1 f_1 = -\zeta$. By the Fredholm alternative, this is solvable if and only if $\zeta \perp \ker(L_1)$, yielding~\eqref{eq:splitting_iff}.
\end{proof}

\emph{On $\mathcal{S}_0$:} every $\varphi \in \ker(L_0^0) \subset \mathcal{S}_0$ is radial, while $\partial_{x_j}\phi \in \mathcal{S}_1$. Since $\mathcal{S}_0 \perp \mathcal{S}_1$, condition~\eqref{eq:splitting_iff} is \emph{automatically satisfied}. Thus eigenvalues can always split from $\Lambda = 0$ on $\mathcal{S}_0$ when $\varepsilon$ crosses an eigenvalue of $\tilde{L}_0^{0}$.

\emph{On $\mathcal{S}_1$:} both $\zeta$ and $\partial_{x_j}\phi$ reside in $\mathcal{S}_1$, so~\eqref{eq:splitting_iff} imposes a genuine constraint.

\begin{proposition}[No splitting from $\lambda_0^{(1)}$]\label{prop:no_splitting_S1}
	Let $\lambda_0^{(1)}$ denote the lowest eigenvalue of $\tilde{L}_0^{1}$, with eigenfunction $\zeta_0(\bx) = \rho_0(r)\,Y_1^m(\hat{\bx})$ where $\rho_0(r) > 0$ for all $r > 0$. Then $\langle\zeta_0, \partial_{x_m}\phi\rangle \neq 0$, and no splitting occurs at $\varepsilon = \lambda_0^{(1)}$.
\end{proposition}

\begin{proof}
	The inner product reduces to $C\int_0^\infty \rho_0(r)\phi'(r)r^{N-1}dr$ with $C > 0$. Since $\rho_0 > 0$ and $\phi' < 0$ for all $r > 0$, the integral is strictly negative, hence non-zero.
\end{proof}

\begin{proposition}[No splitting from $\lambda_1^{(1)}$]\label{prop:no_splitting_S1_excited}
	Let $\lambda_1^{(1)}$ denote the first excited eigenvalue of $\tilde{L}_0^{1}$, with eigenfunction $\zeta_1(\bx) = \rho_1(r)\,Y_1^m(\hat{\bx})$ where $\rho_1(r) > 0$ for all $r \in (0, a)$ and $\rho_1(r) < 0$ for all $r \in (a, \infty)$. Then $\langle\zeta_1, \partial_{x_m}\phi\rangle \neq 0$, and no splitting occurs at $\varepsilon = \lambda_1^{(1)}$.
\end{proposition}

\begin{proof}
	At $\varepsilon = \lambda_1^{(1)}$, $\zeta_1 \in \ker(L_0^1)$ satisfies $L_0\zeta_1 = 0$ (cf.~\eqref{def:tildeL0}), and $\partial_{x_m}\phi \in \ker(L_1)$ by Lemma~\ref{lem:L1_spectral}(ii). Multiplying $L_0\zeta_1 = 0$ by $\partial_{x_m}\phi$, the equation $L_1\partial_{x_m}\phi = 0$ by $\zeta_1$, integrating over $\mathbb{R}^N$ and subtracting gives
	\begin{equation}\label{eq:prop10_identity}
		\varepsilon\int_{\mathbb{R}^N} \zeta_1 \partial_{x_m}\phi\,d\bx = 2\kappa\int_{\mathbb{R}^N}\phi^{2\kappa}\zeta_1 \partial_{x_m} \phi\,d\bx.
	\end{equation}

	We will prove that $\langle \partial_{x_m}\phi, \zeta_1\rangle \neq 0$ by contradiction. If $\langle \partial_{x_m}\phi, \zeta_1\rangle = 0$, then by~\eqref{eq:prop10_identity}, $\langle \partial_{x_m} \phi , \phi^{2\kappa}\zeta_1 \rangle = 0$ as well. Writing these integrals in radial coordinates, both reduce to $C\int_0^\infty \rho_1(r)\phi'(r)\,w(r)\,r^{N-1}\,dr = 0$ with weight $w = 1$ and $w = \phi^{2\kappa}$, respectively. Since $\phi' < 0$ for all $r > 0$, the function $h(r) := -\phi'(r)\rho_1(r)$ changes sign exactly once (from positive to negative at $r = a$). Because $\phi^{2\kappa}$ is a positive, strictly decreasing function, the vanishing of the unweighted integral $\int_0^\infty h\,r^{N-1}\,dr = 0$ together with the Chebyshev integral inequality \citep{hardy:1952} implies $\int_0^\infty \phi^{2\kappa}\,h\,r^{N-1}\,dr > 0$, contradicting $\langle \partial_{x_m}\phi, \phi^{2\kappa}\zeta_1\rangle = 0$.
\end{proof}

\begin{corollary}\label{corolario_bifurcacion}
	Suppose that $\lambda_0^{(1)}$ and $\lambda_1^{(1)}$ exist. Then there exists a unique value of $\varepsilon \in (\lambda_0^{(1)},\lambda_1^{(1)})$ such that the Jordan chain can be extended.
\end{corollary}

\begin{proof}
	By Theorem~\ref{thm:no_splitting}, $\mu(\varepsilon)$ is strictly increasing and diverges at $\lambda_0^{(1)}, \lambda_1^{(1)}$ where $L_0^1$ ceases to be invertible, by Propositions~\ref{prop:no_splitting_S1} and~\ref{prop:no_splitting_S1_excited}. By the intermediate value theorem, there exists a unique $\varepsilon^{\mu} \in (\lambda_0^{(1)}, \lambda_1^{(1)})$ such that $\mu(\varepsilon^{\mu}) = 0$, allowing the Jordan chain to extend.
\end{proof}

\subsection{Direction of splitting in Jordan chains}

In this section, we will determine the behaviour of eigenvalues bifurcating from $\Lambda = 0$ when the Jordan chain extends. The analysis is based on the perturbation theory of linear operators and the structure of the generalised null eigenspace.

\begin{theorem}[Direction of bifurcation on $\mathcal{S}_1$]\label{thm:direction_splitting_S1}
	Under the hypotheses of Lemma~\ref{lem:L1_spectral}, suppose  $\varepsilon^\mu \notin \sigma(\tilde{L}_0^{1})$ is such that $\mu(\varepsilon^\mu) = 0$, where $\mu$ is defined in Theorem~\ref{thm:no_splitting}. Then the Jordan chain extends to length $4N$, and the bifurcating eigenvalues from $\Lambda = 0$ are purely imaginary for $\varepsilon < \varepsilon^\mu$ and real for $\varepsilon > \varepsilon^\mu$.
\end{theorem}

\begin{proof}
	Throughout, we work on the angular momentum subspace $\mathcal{S}_1$, where the eigenvalue problem~\eqref{eq:9kappa}--\eqref{eq:9kappaa} decouples into independent radial problems for each spherical harmonic component $Y_1^m$. By rotational symmetry, the radial operators $L_0^1$ and $L_1^1$ act identically on each component, so it suffices to fix the component $m = 1$. All results extend to $m = 1, \ldots, N$ by symmetry.

	\textit{Part~I: The Jordan chain extends to exactly length~$4$ per component (total $4N$).}

	Define the Jordan chain elements $\eta_k$ in the component $m = 1$ by the recursion $\mathcal{L}\eta_{k+1} = \eta_k$, where $k=1,...$, starting from $\eta_1 = (\partial_{x_1}\phi, 0)^T \in \ker(\mathcal{L})$. Set $f_0 := \partial_{x_1}\phi$ and $\chi := L_0^{-1}(\varepsilon^\mu) f_0$.

	\emph{Length $\geq 2$:} The equation $\mathcal{L}\eta_2 = \eta_1$ requires $L_0 g_2 = f_0$ and $L_1 f_2 = 0$. Since $L_0$ is invertible by hypothesis, $g_2 = \chi$. Choosing $f_2 = 0$, we set $\eta_2 = (0, \chi)^T$.

	\emph{Length $\geq 3$:} The equation $\mathcal{L}\eta_3 = \eta_2$ requires $L_0 g_3 = 0$ and $L_1 f_3 = -\chi$. Invertibility of $L_0$ forces $g_3 = 0$. By the Fredholm alternative, $L_1 f_3 = -\chi$ is solvable if and only if $\chi \perp \ker(L_1)$, i.e., $\langle \partial_{x_k}\phi, \chi \rangle = 0$ for all $k = 1, \ldots, N$. For $k \neq 1$, this holds by angular orthogonality because $\partial_{x_k}\phi$ and $\chi$ belong to different $Y_1^m$ components. For $k = 1$,
	\begin{equation}
		\langle f_0, \chi \rangle = \langle \partial_{x_1}\phi,\, L_0^{-1}(\varepsilon^\mu)\,\partial_{x_1}\phi \rangle = \mu(\varepsilon^\mu) = 0.
	\end{equation}
	Hence, we define $f_3 := -(L_1^1\big|_\perp)^{-1}\chi$, where $L_1^1\big|_\perp$ denotes the restriction of $L_1^1$ to $(\ker L_1^1)^\perp = \Span(f_0)^\perp$. Notice that $f_3$ is well-defined since $\chi \in (\ker L_1^1)^\perp$ and $L_1^1\big|_{\perp} > 0$ by Lemma~\ref{lem:L1_spectral}(ii). Set $\eta_3 = (f_3, 0)^T$.

	\emph{Length $\geq 4$:} The equation $\mathcal{L}\eta_4 = \eta_3$ requires $L_0 g_4 = f_3$ and $L_1 f_4 = 0$. Since $L_0$ is invertible, $g_4 = L_0^{-1}f_3$. Choosing $f_4 = 0$, set $\eta_4 = (0, g_4)^T$.

	\emph{The chain terminates at length~$4$:} For $\mathcal{L}\eta_5 = \eta_4$ to hold, we need $L_0 g_5 = 0$ (forcing $g_5 = 0$) and $L_1 f_5 = -g_4$. By the Fredholm alternative, this requires $\langle f_0, g_4 \rangle = 0$. We calculate
	\begin{equation}
		\langle f_0, g_4 \rangle = \langle f_0,\, L_0^{-1}f_3 \rangle = \langle L_0^{-1}f_0,\, f_3 \rangle = \langle \chi,\, f_3 \rangle = -\langle \chi,\, (L_1^1\big|_\perp)^{-1}\chi \rangle,
	\end{equation}
	using self-adjointness of $L_0^{-1}$. Since $\chi \in (\ker L_1^1)^\perp$ (as shown above) and $\chi \neq 0$ (because $L_0^{-1}$ is invertible and $f_0 \neq 0$), the strict positivity of $(L_1^1\big|_\perp)^{-1}$ gives
	\begin{equation}
		\langle \chi,\, (L_1^1\big|_\perp)^{-1}\chi \rangle > 0.
	\end{equation}
	Therefore $\langle f_0, g_4 \rangle \neq 0$, the Fredholm condition fails, and the chain cannot extend beyond length~$4$.

	Since the chain has length exactly~$4$ independently for each $m = 1, \ldots, N$, the total dimension of the generalised null space of $\mathcal{L}$ restricted to $\mathcal{S}_1$ is~$4N$.

	\medskip
	\textit{Part~II: Direction of bifurcation.}

	The eigenvalue problem on $\mathcal{S}_1$ reduces to
	\begin{equation}\label{eq:reduced_bif}
		L_1 f + \nu\, L_0^{-1}(\varepsilon)\, f = 0, \qquad \nu := \Lambda^2,
	\end{equation}
	for $(\varepsilon, \nu)$ near $(\varepsilon^\mu, 0)$, where $f$ belongs to the component $m = 1$ of $\mathcal{S}_1$. We perform a Lyapunov--Schmidt reduction of~\eqref{eq:reduced_bif} (see, e.g., \citep{kapitula:2013}).

	\emph{Decomposition.} Write $f = f_0 + w$, where $w \perp f_0$ in $L^2$. Project~\eqref{eq:reduced_bif} onto $\operatorname{span}\{f_0\}$ and its orthogonal complement.

	\emph{Auxiliary equation.} The projection onto $\{f_0\}^\perp$ yields
	\begin{equation}\label{eq:aux_eq}
		L_1^1\big|_\perp\, w + \nu\, P^\perp L_0^{-1}(\varepsilon)(f_0 + w) = 0,
	\end{equation}
	where $P^\perp$ denotes the $L^2$-orthogonal projection onto $\{f_0\}^\perp$. Define the map $$G(\nu, \varepsilon, w) := w + \nu\,(L_1^1\big|_\perp)^{-1}\, P^\perp L_0^{-1}(\varepsilon)(f_0 + w).$$

	Then $G(0, \varepsilon, 0) = 0$ and $D_w G(0, \varepsilon, 0) = I$. By the implicit function theorem, there exists a unique smooth solution $w = w(\nu, \varepsilon)$ for $(\nu, \varepsilon)$ near $(0, \varepsilon^\mu)$, satisfying $w(0, \varepsilon) = 0$ for all $\varepsilon$ near $\varepsilon^\mu$, with the expansion
	\begin{equation}\label{eq:w_leading}
		w(\nu, \varepsilon) = -\nu\,(L_1^1\big|_\perp)^{-1}\, P^\perp \chi(\varepsilon) + O(\nu^2),
	\end{equation}
	where $\chi(\varepsilon) := L_0^{-1}(\varepsilon)\, f_0$.

	\emph{Bifurcation equation.} Projecting~\eqref{eq:reduced_bif} onto $\operatorname{span}\{f_0\}$ and using $L_1 f_0 = 0$ gives, for $\nu \neq 0$,
	\begin{equation}\label{eq:bif_eq}
		B(\nu, \varepsilon) := \langle f_0,\, L_0^{-1}(\varepsilon)(f_0 + w(\nu, \varepsilon)) \rangle = 0.
	\end{equation}
	At $(\nu, \varepsilon) = (0, \varepsilon^\mu)$, we have $w = 0$ and $B = \langle f_0,\, L_0^{-1}(\varepsilon^\mu)\, f_0 \rangle = \mu(\varepsilon^\mu) = 0$.

	Now, we verify the hypotheses of the implicit function theorem applied to~\eqref{eq:bif_eq} for $\nu$ as a function of $\varepsilon$.

	\emph{Partial derivative with respect to $\varepsilon$:}  Since $w(0, \varepsilon) = 0$, we have $w_\varepsilon(0, \varepsilon^\mu) = 0$. Using $\frac{d}{d\varepsilon}L_0^{-1} = L_0^{-2}$ (which follows from $L_0' = -I$),
	\begin{equation}\label{eq:B_eps}
		B_\varepsilon(0, \varepsilon^\mu) = \langle f_0,\, L_0^{-2}(\varepsilon^\mu)\, f_0 \rangle = \|L_0^{-1}(\varepsilon^\mu)\, f_0\|^2 = \|\chi(\varepsilon^\mu)\|^2 > 0,
	\end{equation}
	where positivity follows from $\chi(\varepsilon^\mu) \neq 0$ (since $L_0^{-1}$ is invertible and $f_0 \neq 0$).

	\emph{Partial derivative with respect to $\nu$:}  Differentiating~\eqref{eq:bif_eq} gives $B_\nu = \langle f_0,\, L_0^{-1}(\varepsilon)\, w_\nu \rangle$. From~\eqref{eq:w_leading}, $w_\nu(0, \varepsilon^\mu) = -(L_1^1\big|_\perp)^{-1}\, P^\perp \chi(\varepsilon^\mu)$. At $\varepsilon = \varepsilon^\mu$, $\langle f_0, \chi(\varepsilon^\mu) \rangle = 0$ implies $P^\perp \chi(\varepsilon^\mu) = \chi(\varepsilon^\mu)$. Therefore,
	\begin{equation}\label{eq:B_nu}
		B_\nu(0, \varepsilon^\mu) = -\langle \chi(\varepsilon^\mu),\, (L_1^1\big|_\perp)^{-1}\, \chi(\varepsilon^\mu) \rangle < 0,
	\end{equation}
	since $(L_1^1\big|_\perp)^{-1}$ is positive definite.

	\emph{Conclusion.} Since $B_\nu(0, \varepsilon^\mu) \neq 0$, the implicit function theorem guarantees the existence of a unique smooth curve $\varepsilon \mapsto \nu(\varepsilon)$ defined near $\varepsilon^\mu$, with $\nu(\varepsilon^\mu) = 0$ and
	\begin{equation}\label{eq:nu_deriv}
		\nu'(\varepsilon^\mu) = -\frac{B_\varepsilon(0, \varepsilon^\mu)}{B_\nu(0, \varepsilon^\mu)} = \frac{\|\chi(\varepsilon^\mu)\|^2}{\langle \chi(\varepsilon^\mu),\, (L_1^1\big|_\perp)^{-1}\, \chi(\varepsilon^\mu) \rangle} > 0.
	\end{equation}

	Since $\nu = \Lambda^2$ and $\nu'(\varepsilon^\mu) > 0$, for $\varepsilon$ sufficiently close to $\varepsilon^\mu$:
	\begin{itemize}
		\item If $\varepsilon < \varepsilon^\mu$, then $\nu(\varepsilon) < 0$, so $\Lambda^2 < 0$ and the bifurcating eigenvalues are purely imaginary: $\Lambda = \pm i\sqrt{|\nu(\varepsilon)|}$.
		\item If $\varepsilon > \varepsilon^\mu$, then $\nu(\varepsilon) > 0$, so $\Lambda^2 > 0$ and the bifurcating eigenvalues are real: $\Lambda = \pm\sqrt{\nu(\varepsilon)}$.
	\end{itemize}
\end{proof}

\begin{theorem}[Bifurcation at eigenvalues of $\tilde{L}_0^{1}$]\label{thm:bifurcation_L0_eigenvalue}
	Under the hypotheses of Lemma~\ref{lem:L1_spectral}, suppose  $\varepsilon^\mu \in \sigma(\tilde{L}_0^{1})$ is a simple eigenvalue, with eigenfunction $\zeta \in \mathcal{S}_1$ satisfying $\tilde{L}_0^{1}\zeta = \varepsilon^\mu\zeta$. Assume $\langle \zeta, \partial_{x_j}\phi \rangle = 0$ for all $j = 1, \ldots, N$, and the non-degeneracy condition
	\begin{equation}\label{eq:nondegen_L0}
		\langle \partial_{x_1}\phi,\, (L_0^1\big|_{\{\zeta\}^\perp})^{-1}\partial_{x_1}\phi\rangle \neq 0.
	\end{equation}
	Then the generalised null space of $\mathcal{L}(\varepsilon^\mu)$ restricted to $\mathcal{S}_1$ has dimension $4N$, consisting of $2N$ Jordan blocks of size~$2$. Moreover, for $\varepsilon$ near $\varepsilon^\mu$, the eigenvalues bifurcating from $\Lambda = 0$ satisfy
	\begin{equation}\label{eq:bif_L0_eigenvalue}
		\Lambda^2 = \sigma_0\,(\varepsilon - \varepsilon^\mu) + O\!\left((\varepsilon - \varepsilon^\mu)^2\right), \qquad \sigma_0 := \frac{\|\zeta\|^2}{\langle \zeta,\,(L_1^1\big|_\perp)^{-1}\zeta\rangle} > 0,
	\end{equation}
	so the bifurcating eigenvalues are purely imaginary for $\varepsilon < \varepsilon^\mu$ and real for $\varepsilon > \varepsilon^\mu$.
\end{theorem}

\begin{proof}
	By rotational symmetry, it suffices to work on the $j = 1$ component of $\mathcal{S}_1$. Set $f_0 := \partial_{x_1}\phi$. As in the proof of Theorem~\ref{thm:direction_splitting_S1}, $L_1$ does not depend on $\varepsilon$ and $\frac{dL_0}{d\varepsilon} = -I$.

	\medskip
	\textit{Part~I: Structure of the generalised null space at $\varepsilon = \varepsilon^\mu$.}

	At $\varepsilon = \varepsilon^\mu$, $L_0(\varepsilon^\mu)$ restricted to $\mathcal{S}_1$ has $\ker(L_0^1) = \operatorname{span}\{\zeta\}$ (since $\varepsilon^\mu$ is a simple eigenvalue of $\tilde{L}_0^{1}$).

	\emph{Kernel of $\mathcal{L}$:} $\mathcal{L}(f,g)^T = 0$ requires $L_0 g = 0$ and $L_1 f = 0$. Hence $g \in \operatorname{span}\{\zeta\}$ and $f \in \operatorname{span}\{f_0\}$ (by Lemma~\ref{lem:L1_spectral}(ii)), so $\ker(\mathcal{L})= \operatorname{span}\{e_1, e_3\}$ with
	\begin{equation}
		e_1 := (f_0, 0)^T, \qquad e_3 := (0, \zeta)^T.
	\end{equation}

	\emph{Jordan chain from $e_1$:} We seek $e_2$ such that $\mathcal{L}e_2 = e_1$, i.e., $L_0 g_2 = f_0$ and $L_1 f_2 = 0$. The Fredholm alternative requires $f_0 \perp \ker(L_0^1) = \operatorname{span}\{\zeta\}$, which is guaranteed by the hypothesis $\langle \zeta, f_0 \rangle = 0$. Let $g_2 := (L_0^1\big|_{\zeta^\perp})^{-1}f_0$ be the unique solution in $\{\zeta\}^\perp$. Choosing $f_2 = 0$, we define
	\begin{equation}
		e_2 := (0,\, g_2)^T.
	\end{equation}

	\emph{Jordan chain from $e_3$:} We seek $e_4$ such that $\mathcal{L}e_4 = e_3$, i.e., $L_0 g_4 = 0$ and $L_1 f_4 = -\zeta$. From the first equation $g_4 \in \operatorname{span}\{\zeta\}$; choose $g_4 = 0$. For the second, the Fredholm alternative requires $\zeta \perp \ker(L_1^1) = \operatorname{span}\{f_0\}$, which holds by hypothesis. Since $L_1^1\big|_{\perp} > 0$ by Lemma~\ref{lem:L1_spectral}(ii), define
	\begin{equation}
		e_4 := \bigl(-(L_1^1\big|_\perp)^{-1}\zeta,\, 0\bigr)^T.
	\end{equation}

	\emph{No further extension from $e_1$:} To extend, we need $\mathcal{L}\xi = e_2$, i.e., $L_0 g_\xi = 0$ and $L_1 f_\xi = -g_2$. The Fredholm condition for $L_0^0$ requires $\langle f_0, g_2 \rangle = 0$. But $g_2 = (L_0^1\big|_{\zeta^\perp})^{-1}f_0$ and
	\begin{equation}
		\langle f_0, g_2 \rangle = \langle f_0,\, (L_0^1\big|_{\zeta^\perp})^{-1} f_0 \rangle \neq 0
	\end{equation}
	by the non-degeneracy hypothesis~\eqref{eq:nondegen_L0}. Hence the chain from $e_1$ terminates at length~$2$.

	\emph{No further extension from $e_3$:} To extend, we need $\mathcal{L}\xi = e_4$, i.e., $L_0 g_\xi = -(L_1^1\big|_\perp)^{-1}\zeta$ and $L_1 f_\xi = 0$. The Fredholm condition for $L_0 g_\xi = -(L_1^1\big|_\perp)^{-1}\zeta$ requires
	\begin{equation}
		\langle \zeta,\, (L_1^1\big|_\perp)^{-1}\zeta \rangle = 0.
	\end{equation}
	Since $(L_1^1\big|_\perp)^{-1}$ is positive definite on $(\ker L_1^1)^\perp$ and $\zeta \in (\ker L_1^1)^\perp$ with $\zeta \neq 0$, this inner product is strictly positive. The Fredholm condition fails, and the chain from $e_3$ terminates at length~$2$.

	Therefore $\ker(\mathcal{L}^2) = \operatorname{span}\{e_1, e_2, e_3, e_4\}$ has dimension~$4$ per directional component, and the generalised null space of $\mathcal{L}(\varepsilon^\mu)$ restricted to $\mathcal{S}_1$ has total dimension~$4N$.

	\medskip
	\textit{Part~II: Direction of bifurcation via the resolvent expansion.}

	For $\varepsilon \neq \varepsilon^\mu$ sufficiently close to $\varepsilon^\mu$, $L_0(\varepsilon)$ is invertible on $\mathcal{S}_1$ (since $\varepsilon^\mu$ is an isolated simple eigenvalue of $\tilde{L}_0^{1}$), and the eigenvalue problem on $\mathcal{S}_1$ reduces to
	\begin{equation}\label{eq:reduced_ev_L0}
		-L_1 f = \nu\, L_0^{-1}(\varepsilon)\, f, \qquad \nu := \Lambda^2.
	\end{equation}

	\emph{Step~1: Resolvent expansion.}
	Since $\tilde{L}_0^{1}\zeta = \varepsilon^\mu\zeta$ is a simple eigenvalue, $L_0(\varepsilon)\zeta = (\varepsilon^\mu - \varepsilon)\zeta$, whence (see, e.g., \citep[Chapter~III]{kato:1995})
	\begin{equation}\label{eq:resolvent_expansion}
		L_0^{-1}(\varepsilon) = \frac{P_\zeta}{\varepsilon^\mu - \varepsilon} + R(\varepsilon),
	\end{equation}
	where $P_\zeta := \frac{|\zeta\rangle\langle\zeta|}{\|\zeta\|^2}$ is the orthogonal projection onto $\operatorname{span}\{\zeta\}$, and
	\begin{equation}
		R(\varepsilon) := \bigl(L_0(\varepsilon)\big|_{\{\zeta\}^\perp}\bigr)^{-1}(I - P_\zeta)
	\end{equation}
	is bounded and analytic in $\varepsilon$ near $\varepsilon^\mu$.

	\emph{Step~2: Rescaling.}
	Substituting~\eqref{eq:resolvent_expansion} into~\eqref{eq:reduced_ev_L0} gives
	\begin{equation}
		-L_1 f = \frac{\nu}{\varepsilon^\mu - \varepsilon}\, P_\zeta f + \nu\, R(\varepsilon)\, f.
	\end{equation}
	Setting $\nu = (\varepsilon - \varepsilon^\mu)\,\sigma$ (where $\sigma$ remains to be determined), we obtain
	\begin{equation}\label{eq:rescaled_ev}
		-L_1 f + \sigma\, P_\zeta f = (\varepsilon - \varepsilon^\mu)\,\sigma\, R(\varepsilon)\, f.
	\end{equation}
	For $|\varepsilon - \varepsilon^\mu|$ small, the right-hand side is a perturbation. The limiting problem at $\varepsilon = \varepsilon^\mu$ reads
	\begin{equation}\label{eq:limiting_problem}
		L_1 f = \sigma\, P_\zeta f = \frac{\sigma\, \langle\zeta, f\rangle}{\|\zeta\|^2}\,\zeta.
	\end{equation}

	\emph{Step~3: Solution of the limiting problem.}
	We solve~\eqref{eq:limiting_problem} subject to the normalisation $\langle\zeta, f\rangle = 1$. The equation becomes $L_1 f = \frac{\sigma}{\|\zeta\|^2}\,\zeta$. By the Fredholm alternative (applied with $\ker(L_1^1) = \operatorname{span}\{f_0\}$ and $\langle f_0, \zeta\rangle = 0$), the unique solution in $(\ker L_1^1)^\perp$ is
	\begin{equation}
		h_0 = \frac{\sigma}{\|\zeta\|^2}\,(L_1^1\big|_\perp)^{-1}\zeta.
	\end{equation}
	The general solution is $f = h_0 + c\,f_0$ for any $c \in \mathbb{R}$. Imposing $\langle\zeta, f\rangle = 1$ and using $\langle\zeta, f_0\rangle = 0$:
	\begin{equation}
		1 = \langle\zeta, h_0\rangle = \frac{\sigma}{\|\zeta\|^2}\,\langle\zeta,\,(L_1^1\big|_\perp)^{-1}\zeta\rangle.
	\end{equation}
	Solving for $\sigma$ gives
	\begin{equation}\label{eq:sigma_0}
		\sigma_0 = \frac{\|\zeta\|^2}{\langle\zeta,\,(L_1^1\big|_\perp)^{-1}\zeta\rangle} > 0,
	\end{equation}
	where positivity follows because $(L_1^1\big|_\perp)^{-1}$ is positive definite and $\zeta \in (\ker L_1^1)^\perp \setminus\{0\}$.

	\emph{Step~4: Implicit function theorem.}
	Decompose $f = h + c\,f_0$ with $h \in V := (\ker L_1^1)^\perp \cap \mathcal{S}_1$ and $c \in \mathbb{R}$. Since $\langle f_0, \zeta\rangle = 0$, the projection of~\eqref{eq:rescaled_ev} onto $\ker(L_1^1) = \operatorname{span}\{f_0\}$ yields
	\begin{equation}
		0 = (\varepsilon - \varepsilon^\mu)\,\sigma\,\langle f_0,\, R(\varepsilon)(h + c\,f_0)\rangle.
	\end{equation}
	For $\varepsilon \neq \varepsilon^\mu$ and $\sigma \neq 0$, this determines $c$ uniquely:
	\begin{equation}\label{eq:c_equation}
		c(\varepsilon, h) = -\frac{\langle f_0,\, R(\varepsilon)\,h\rangle}{\langle f_0,\, R(\varepsilon)\,f_0\rangle},
	\end{equation}
	which is well-defined and smooth near $(\varepsilon^\mu, h_0)$ by the non-degeneracy hypothesis~\eqref{eq:nondegen_L0}, since $R(\varepsilon^\mu)f_0 = (L_0^1\big|_{\zeta^\perp})^{-1}f_0$ and $\langle f_0, R(\varepsilon^\mu)f_0\rangle \neq 0$.

	The projection of~\eqref{eq:rescaled_ev} onto $V$, with the normalisation $\langle\zeta, h\rangle = 1$ (which is consistent since $\langle\zeta, f_0\rangle = 0$), gives
	\begin{equation}\label{eq:V_projection}
		\Psi(\varepsilon, \sigma, h) := L_1^1\big|_\perp\, h - \frac{\sigma}{\|\zeta\|^2}\,\zeta + (\varepsilon - \varepsilon^\mu)\,\sigma\,P^\perp R(\varepsilon)\bigl(h + c(\varepsilon, h)\,f_0\bigr) = 0,
	\end{equation}
	where $P^\perp$ projects onto $V$. At $(\varepsilon^\mu, \sigma_0, h_0)$ with $h_0 = \frac{\sigma_0}{\|\zeta\|^2}(L_1^1\big|_\perp)^{-1}\zeta$, we have $\Psi = 0$.

	We apply the implicit function theorem to solve for $(\sigma, h)$ jointly as a function of $\varepsilon$. The tangent space to the constraint $\langle\zeta, h\rangle = 1$ is $W := \{v \in V : \langle\zeta, v\rangle = 0\}$. At $(\varepsilon^\mu, \sigma_0, h_0)$, the partial derivatives are:
	\begin{equation}
		D_h\Psi\cdot\delta h = L_1^1\big|_\perp\,\delta h, \quad D_\sigma\Psi = -\frac{\zeta}{\|\zeta\|^2},
	\end{equation}
	for $\delta h \in W$ (note that $\langle\zeta, \delta h\rangle = 0$ eliminates the $P_\zeta$ contribution, and the $(\varepsilon - \varepsilon^\mu)$ term vanishes at $\varepsilon = \varepsilon^\mu$). The joint Fr\'echet derivative $D_{(\sigma, h)}\Psi \colon \mathbb{R} \times W \to V$ acts as
	\begin{equation}
		D_{(\sigma, h)}\Psi\cdot(\delta\sigma, \delta h) = L_1^1\big|_\perp\,\delta h - \frac{\delta\sigma}{\|\zeta\|^2}\,\zeta.
	\end{equation}
	We verify this is a bijection. Given any $v \in V$, we seek $(\delta\sigma, \delta h) \in \mathbb{R} \times W$ such that $L_1^1\big|_\perp\,\delta h = v + \frac{\delta\sigma}{\|\zeta\|^2}\,\zeta$. Applying $(L_1^1\big|_\perp)^{-1}$ gives
	\begin{equation}
		\delta h = (L_1^1\big|_\perp)^{-1}\!\left(v + \frac{\delta\sigma}{\|\zeta\|^2}\,\zeta\right).
	\end{equation}
	The constraint $\langle\zeta, \delta h\rangle = 0$ then uniquely determines
	\begin{equation}
		\delta\sigma = -\frac{\|\zeta\|^2\,\langle\zeta,\,(L_1^1\big|_\perp)^{-1}v\rangle}{\langle\zeta,\,(L_1^1\big|_\perp)^{-1}\zeta\rangle},
	\end{equation}
	which is well-defined since $\langle\zeta,\,(L_1^1\big|_\perp)^{-1}\zeta\rangle = \|\zeta\|^2/\sigma_0 > 0$. This shows $D_{(\sigma,h)}\Psi$ is a bijection. By the implicit function theorem, for $\varepsilon$ near $\varepsilon^\mu$ there exist unique smooth functions $\sigma(\varepsilon)$ and $h(\varepsilon)$ with $\sigma(\varepsilon^\mu) = \sigma_0$ and $h(\varepsilon^\mu) = h_0$.

	\emph{Step~5: Conclusion.}
	Since $\nu = \Lambda^2 = (\varepsilon - \varepsilon^\mu)\,\sigma(\varepsilon)$ and $\sigma(\varepsilon^\mu) = \sigma_0 > 0$, for $\varepsilon$ sufficiently close to $\varepsilon^\mu$:
	\begin{itemize}
		\item If $\varepsilon < \varepsilon^\mu$, then $\Lambda^2 = (\varepsilon - \varepsilon^\mu)\sigma(\varepsilon) < 0$, so $\Lambda = \pm i\sqrt{|\Lambda^2|}$ is purely imaginary.
		\item If $\varepsilon > \varepsilon^\mu$, then $\Lambda^2 = (\varepsilon - \varepsilon^\mu)\sigma(\varepsilon) > 0$, so $\Lambda = \pm\sqrt{\Lambda^2}$ is real.
	\end{itemize}
\end{proof}

\subsection{Degenerate bifurcation on \texorpdfstring{$\mathcal{S}_1$}{S1}}\label{sec:degenerate_S1}

We now consider the case in which the non-degeneracy condition~\eqref{eq:nondegen_L0} fails at an eigenvalue $\varepsilon^\mu \in \sigma(\tilde{L}_0^{1})$, i.e.,
\begin{equation}\label{eq:degen_condition_S1}
	\langle \partial_{x_i}\phi,\, (L_0^1\big|_{\{\zeta\}^\perp})^{-1}\partial_{x_i}\phi\rangle = 0, \qquad \text{for } i \in \{1, \ldots, N\}.
\end{equation}
When this holds, the Jordan chain from $e_1 = (f_0, 0)^T$ extends beyond length~$2N$, while the chain from $e_3 = (0, \zeta)^T$ with $\ker(L_0^1) = \operatorname{span}\{\zeta\}$ remains of length~$2N$ (since $\langle \zeta, (L_1^1\big|_\perp)^{-1}\zeta\rangle > 0$ always).

\begin{theorem}[Degenerate bifurcation at eigenvalues of $\tilde{L}_0^{1}$]\label{thm:degenerate_S1}
	Under the hypotheses of Theorem~\ref{thm:bifurcation_L0_eigenvalue}, suppose the degeneracy condition~\eqref{eq:degen_condition_S1} holds instead of~\eqref{eq:nondegen_L0}. Then:
	\begin{enumerate}[label=(\roman*)]
		\item The generalised null space of $\mathcal{L}(\varepsilon^\mu)$ restricted to $\mathcal{S}_1$ has dimension~$6N$, consisting of $N$ Jordan blocks of size~$4$ and $N$ blocks of size~$2$.

		\item As $\varepsilon$ increases through~$\varepsilon^\mu$, the number of real eigenvalues on~$\mathcal{S}_1$ either:
		      \begin{itemize}
			      \item increases by~$4N$, if the eigenvalues from the $e_1$ chain also bifurcate from the imaginary axis to the real axis;
			      \item increases by~$2N$, if the eigenvalues from the $e_1$ chain remain on the same initial axis on both sides of~$\varepsilon^\mu$; or
			      \item remains unchanged, if the eigenvalues from the $e_1$ chain pass from the real axis to the imaginary axis, compensating the gain from the $e_3$ chain.
		      \end{itemize}
	\end{enumerate}
\end{theorem}

\begin{proof}
	By rotational symmetry, it suffices to work on the $j = 1$ component of $\mathcal{S}_1$. Set $f_0 := \partial_{x_1}\phi$. As before, $L_1$ does not depend on $\varepsilon$ and $\frac{dL_0}{d\varepsilon} = -I$.

	\medskip
	\textit{Part~I: The Jordan chain from $e_1$ extends to length~$4$.}

	Recall from the proof of Theorem~\ref{thm:bifurcation_L0_eigenvalue}:
	\begin{equation}
		e_1 = (f_0, 0)^T, \quad e_2 = (0,\, g_2)^T, \quad g_2 = (L_0^1\big|_{\zeta^\perp})^{-1}f_0,
	\end{equation}
	which satisfies $\mathcal{L}e_2 = e_1$.

	\emph{Extension to length~$3$:} The equation $\mathcal{L}e_3' = e_2$ requires $L_0^1 g_3' = 0$ and $L_1^1 f_3' = -g_2$. The first gives $g_3' \in \operatorname{span}\{\zeta\}$; choose $g_3' = 0$. The Fredholm condition for $L_1^1 f_3' = -g_2$ requires $\langle f_0, g_2 \rangle = 0$, which is precisely the degeneracy condition~\eqref{eq:degen_condition_S1}. Since $g_2 \perp f_0 \in \ker(L_1^1)$, the unique solution in $(\ker L_1^1)^\perp$ is
	\begin{equation}
		f_3' = -(L_1^1\big|_\perp)^{-1}g_2 = -h_3.
	\end{equation}
	Set $e_3' := (-h_3, 0)^T$.

	\emph{Extension to length~$4$:} The equation $\mathcal{L}e_4' = e_3'$ requires $L_0^1 g_4' = -h_3$ and $L_1^1 f_4' = 0$. The Fredholm condition for $L_0^1 g_4' = -h_3$ requires $\langle \zeta, h_3 \rangle = 0$, which holds by symmetry considerations of the spectra. Since $h_3 \in \{\zeta\}^\perp$, define $g_4' := -(L_0^1\big|_{\zeta^\perp})^{-1}h_3 =: -h_4$ and set
	\begin{equation}
		e_4' := (0,\, -h_4)^T.
	\end{equation}

	\emph{The chain terminates at length~$4$:} For $\mathcal{L}\xi = e_4'$, we need $L_0^1 g_\xi = 0$ (thus $g_\xi \in \operatorname{span}\{\zeta\}$) and $L_1^1 f_\xi = h_4$. The Fredholm condition cannot be fulfilled because
	\begin{equation}
		\langle f_0,\, h_4 \rangle = \langle g_2,\, h_3 \rangle = \langle g_2,\, (L_1^1\big|_\perp)^{-1}\, g_2 \rangle > 0,
	\end{equation}
	where the first equality follows from self-adjointness of $(L_0^1\big|_{\zeta^\perp})^{-1}$ on $\{\zeta\}^\perp$ (noting $f_0, h_3 \in \{\zeta\}^\perp$), and positivity from the positive definiteness of $(L_1^1\big|_\perp)^{-1}$.

	Hence the chain from $e_1$ terminates at exactly length~$4$.

	The chain from $e_3 = (0, \zeta)^T$ is unaffected: as shown in Theorem~\ref{thm:bifurcation_L0_eigenvalue}, the extension requires $\langle \zeta, (L_1^1\big|_\perp)^{-1}\zeta\rangle = 0$, which fails since $(L_1^1\big|_\perp)^{-1}$ is positive definite. Hence the $e_3$ chain has length~$2$.

	The total generalised null space per component has dimension~$4 + 2 = 6$, giving dimension $6N$ on $\mathcal{S}_1$.

	\medskip
	\textit{Part~II: Bifurcation from the $e_3$ chain and the $e_1$ chain.}

	The $e_3$ chain is independent of the $e_1$ chain, and its bifurcation is identical to the non-degenerate case analysed in Theorem~\ref{thm:bifurcation_L0_eigenvalue}: the eigenvalues satisfy $\Lambda^2 = \sigma_0\,(\varepsilon - \varepsilon^\mu) + O((\varepsilon - \varepsilon^\mu)^2)$ with $\sigma_0 > 0$, so $2N$ eigenvalues pass from the imaginary axis to the real axis as $\varepsilon$ increases through~$\varepsilon^\mu$.

	For the $e_1$ chain, the Hamiltonian structure forces all eigenvalues to satisfy $\Lambda \mapsto -\Lambda$ and $\Lambda \mapsto \Lambda^\star$. In particular, eigenvalues that are not at the origin must lie on either the real or the imaginary axis, due to Proposition~\ref{prop:l1_structure}. By continuity, for $\varepsilon$ near~$\varepsilon^\mu$ the $2N$ eigenvalues bifurcating from this chain remain on one of these two axes on each side of~$\varepsilon^\mu$, yielding the three scenarios described in  (ii).
\end{proof}

The results regarding this section allow us to deduce that the following instability criterion for $\mathcal{S}_1$ is sharp:

\begin{corollary}[Instability criterion on $\mathcal{S}_1$]\label{cor:instability_S1}
	Let $\mu(\varepsilon) = \langle \partial_{x_1}\phi,\, (L_0^1)^{-1}\partial_{x_1}\phi\rangle$ be as in Theorem~\ref{thm:no_splitting}, and let $\lambda_0^{(1)} < \lambda_1^{(1)}$ denote the first two eigenvalues of $\tilde{L}_0^{1}$.
	\begin{enumerate}[label=(\roman*)]
		\item If $\mu(\varepsilon) \neq 0$ for every $\varepsilon \in [0,1]$, then the eigenvalue problem~\eqref{eq:9kappa}--\eqref{eq:9kappaa} restricted to $\mathcal{S}_1$ has no eigenvalue with positive real part for any $\varepsilon \in [0,1]$.
		\item If $\mu(\varepsilon) = 0$ for some $\varepsilon \in [0,1]$, then there exists a unique $\varepsilon^\mu \in (\lambda_0^{(1)},\, \min(1, \lambda_1^{(1)}))$ (with the convention $\min(1,\lambda_1^{(1)}) = 1$ when $\lambda_1^{(1)}$ does not exist). Moreover, the eigenvalue problem on $\mathcal{S}_1$ satisfies:
		      \begin{itemize}
			      \item for $\varepsilon \leq \varepsilon^\mu$, all eigenvalues are purely imaginary;
			      \item for $\varepsilon > \varepsilon^\mu$, there always exists an eigenvalue with positive real part.
		      \end{itemize}
	\end{enumerate}
\end{corollary}

\begin{proof}
	Part~(i) follows from Theorem~\ref{thm:no_splitting}: when $\mu(\varepsilon) \neq 0$, the generalised null space of $\mathcal{L}$ on $\mathcal{S}_1$ has dimension $2N$ and no Jordan chain extends, so no eigenvalue can leave the imaginary axis.

	For part~(ii), Corollary~\ref{corolario_bifurcacion} ensures the existence and uniqueness of $\varepsilon^\mu \in (\lambda_0^{(1)}, \lambda_1^{(1)})$. Theorems~\ref{thm:direction_splitting_S1} and~\ref{thm:degenerate_S1} then show that $2N$ eigenvalues bifurcate from $\Lambda = 0$ into the real axis for $\varepsilon > \varepsilon^\mu$, while they remain purely imaginary for $\varepsilon < \varepsilon^\mu$. Since the bifurcation produces real eigenvalues that persist for $\varepsilon > \varepsilon^\mu$ by continuity, instability holds for all $\varepsilon > \varepsilon^\mu$.
\end{proof}

\subsection{Bifurcation on \texorpdfstring{$\mathcal{S}_0$}{S0} at eigenvalues of \texorpdfstring{$\tilde{L}_0^{0}$}{L0 tilde in S0}}

The splitting analysis on the radial sector $\mathcal{S}_0$ differs fundamentally from the $\mathcal{S}_1$ case studied in Theorems~\ref{thm:direction_splitting_S1}--\ref{thm:degenerate_S1}: the Fredholm orthogonality condition~\eqref{eq:splitting_iff} is automatically satisfied (since $\mathcal{S}_0 \perp \mathcal{S}_1$), but the operator $L_1^0$ has exactly one negative eigenvalue (Lemma~\ref{lem:L1_spectral}(i)), so the sign of the bifurcation is not determined a priori.

Let $\{\lambda_n^{(0)}\}_{n \geq 0}$ denote the eigenvalues of $\tilde{L}_0^{0}$ in increasing order, with $\lambda_0^{(0)} = 0$ (Lemma~\ref{lem:L0_spectral}).

\begin{lemma}\label{condicion_no_degeneracion}
	Suppose $\varepsilon^* = \lambda_n^{(0)}$ for some $n \geq 1$, and that $\lambda_n^{(0)}$ is a simple eigenvalue of $\tilde{L}_0^{0}$ with eigenfunction $\varphi \in \mathcal{S}_0$. Then
	\begin{equation}\label{eq:nondegen_S0}
		\langle \varphi,\, (L_1^0)^{-1}\varphi \rangle > 0.
	\end{equation}
\end{lemma}

\begin{proof}
	For convenience, we write $f = (L_1^0)^{-1} \varphi$ and must show that
	\begin{equation}\label{eq:nondegen_S03}
		\langle f,\, L_1^0 f \rangle > 0.
	\end{equation}

	Consider the projection of the equation $L_1 f = \varphi$ onto $\phi$, noting that $\tilde{L}_0 \phi = 0$.
	This yields the relationship:
	\begin{equation}
		\langle \phi,\, L_1 f \rangle = -2\kappa\langle \phi^{2\kappa+1},\,f \rangle = \langle \phi,\, \varphi \rangle = 0,
	\end{equation}
	due to the orthogonality between eigenfunctions associated with different eigenvalues of a Hermitian operator.

	According to Lemma~\ref{lem:L1_spectral}(v), there exists $C>0$ such that
	\begin{equation}
		\langle f,\, L_1^0 f \rangle \geq C \langle \phi, f\rangle^2,
	\end{equation}
	because $f \in \mathcal{S}_0$ and $f \in \{\phi^{2\kappa+1}\}^\perp$ by the previous calculation.

	We now prove that $\langle f,\, L_1^0 f \rangle > 0$. Suppose by contradiction that $\langle f,\, L_1^0 f \rangle = 0$. Then, by the previous inequality, we have $\langle \phi, f\rangle = 0$. Hence $f$ achieves the minimum value $0$ of $\Sigma(h) := \langle h,\, L_1^0 h \rangle$ over the subspace $\{h \in \mathcal{S}_0 : \langle \phi, h\rangle = 0, \langle \phi^{2\kappa+1}, h\rangle = 0\}$, so it is a critical point there. By the Lagrange multipliers theorem, there exist $\alpha, \beta \in \mathbb{R}$ such that
	\begin{equation}
		L_1^0 f = \alpha \phi + \beta \phi^{2\kappa+1}.
	\end{equation}
	By the definition of $f$, we also have $L_1^0 f = \varphi$. Hence $\alpha \phi + \beta \phi^{2\kappa+1} = \varphi$, obtaining a contradiction due to the different exponential decay of $\phi$ and $\varphi$.
	Thus $\langle f,\, L_1^0 f \rangle > 0$.
\end{proof}

\begin{theorem}[Bifurcation at eigenvalues of $\tilde{L}_0^{0}$]\label{thm:bifurcation_S0}
	Under the hypotheses of Lemma~\ref{lem:L1_spectral}, suppose  $\varepsilon^* = \lambda_n^{(0)}$ for some $n \geq 1$, and that $\lambda_n^{(0)}$ is a simple eigenvalue of $\tilde{L}_0^{0}$ with eigenfunction $\varphi \in \mathcal{S}_0$.

	Then the generalised null space of $\mathcal{L}(\varepsilon^*)$ restricted to $\mathcal{S}_0$ admits a Jordan block of size~$2$ associated with $\varphi$. Moreover, for $\varepsilon$ near $\varepsilon^*$, the eigenvalues bifurcating from $\Lambda = 0$ satisfy
	\begin{equation}\label{eq:bif_S0}
		\Lambda^2 = \sigma_0^{(0)}\,(\varepsilon - \varepsilon^*) + O\!\left((\varepsilon - \varepsilon^*)^2\right), \qquad \sigma_0^{(0)} := \frac{\|\varphi\|^2}{\langle \varphi,\, (L_1^0)^{-1}\varphi\rangle} >0,
	\end{equation}
	where $L_1^0$ is invertible on $\mathcal{S}_0$ (since $\ker(L_1) \subset \mathcal{S}_1$). Then eigenvalues are purely imaginary for $\varepsilon < \varepsilon^*$ and real for $\varepsilon > \varepsilon^*$.
\end{theorem}

\begin{proof}
	The proof follows the same resolvent expansion strategy as in Theorem~\ref{thm:bifurcation_L0_eigenvalue}, adapted to the radial sector $\mathcal{S}_0$ where the operator $L_1^0$ is invertible but indefinite.

	\medskip
	\textit{Part~I: Structure of the generalised null space at $\varepsilon = \varepsilon^*$.}

	At $\varepsilon = \varepsilon^*$, $L_0^0(\varepsilon^*) = \tilde{L}_0^{0} - \varepsilon^*$ has $\ker(L_0^0) = \operatorname{span}\{\varphi\}$ (since $\varepsilon^*$ is a simple eigenvalue of $\tilde{L}_0^{0}$). The kernel $\ker(L_1^0) = \{0\}$ on $\mathcal{S}_0$, since $\ker(L_1) \subset \mathcal{S}_1$ by Lemma~\ref{lem:L1_spectral}(ii).

	\emph{Kernel of $\mathcal{L}$ on $\mathcal{S}_0$:} $\mathcal{L}(f,g)^T = 0$ requires $L_0 g = 0$ and $L_1 f = 0$. On $\mathcal{S}_0$, $L_1^0$ is invertible, so $f = 0$. Hence $\ker(\mathcal{L}\big|_{\mathcal{S}_0}) = \operatorname{span}\{e_1\}$ with
	\begin{equation}
		e_1 := (0, \varphi)^T.
	\end{equation}

	\emph{Jordan chain from $e_1$:} We seek $e_2$ such that $\mathcal{L}e_2 = e_1$, i.e., $L_0 g_2 = 0$ and $L_1 f_2 = -\varphi$. The first equation gives $g_2 \in \operatorname{span}\{\varphi\}$; choose $g_2 = 0$. For the second, since $L_1^0$ is invertible on $\mathcal{S}_0$, the solution exists uniquely:
	\begin{equation}
		e_2 := \bigl(-(L_1^0)^{-1}\varphi,\, 0\bigr)^T.
	\end{equation}

	\emph{The chain terminates at length~$2$:} For $\mathcal{L}\xi = e_2$ to hold, we need $L_0 g_\xi = -(L_1^0)^{-1}\varphi$ and $L_1 f_\xi = 0$ (so $f_\xi = 0$). The Fredholm condition for $L_0^0$ requires $\langle \varphi, (L_1^0)^{-1}\varphi \rangle = 0$, which is excluded by the non-degeneracy condition of Lemma~\ref{condicion_no_degeneracion}. Hence the chain terminates at length~$2$.

	\medskip
	\textit{Part~II: Direction of bifurcation via the resolvent expansion.}

	For $\varepsilon \neq \varepsilon^*$ sufficiently close to $\varepsilon^*$, $L_0^0(\varepsilon)$ is invertible on $\mathcal{S}_0$ and we reduce the eigenvalue problem on $\mathcal{S}_0$ to
	\begin{equation}\label{eq:reduced_ev_S0}
		-L_1^0 f = \nu\, \bigl(L_0^0(\varepsilon)\bigr)^{-1} f, \qquad \nu := \Lambda^2.
	\end{equation}

	\emph{Step~1: Resolvent expansion.}
	Since $L_0^0(\varepsilon)\varphi = (\varepsilon^* - \varepsilon)\varphi$, the resolvent admits the Laurent expansion (cf. \citep[Chapter~III]{kato:1995})
	\begin{equation}\label{eq:resolvent_S0}
		\bigl(L_0^0(\varepsilon)\bigr)^{-1} = \frac{P_\varphi}{\varepsilon^* - \varepsilon} + R_0(\varepsilon),
	\end{equation}
	where $P_\varphi := \frac{|\varphi\rangle\langle\varphi|}{\|\varphi\|^2}$ and $R_0(\varepsilon) := \bigl(L_0^0(\varepsilon)\big|_{\{\varphi\}^\perp \cap \mathcal{S}_0}\bigr)^{-1}(I - P_\varphi)$ is bounded and analytic near $\varepsilon^*$.

	\emph{Step~2: Rescaling.}
	Setting $\nu = (\varepsilon - \varepsilon^*)\,\sigma$ in~\eqref{eq:reduced_ev_S0} and substituting~\eqref{eq:resolvent_S0}:
	\begin{equation}\label{eq:rescaled_S0}
		-L_1^0 f + \sigma\, P_\varphi f = (\varepsilon - \varepsilon^*)\,\sigma\, R_0(\varepsilon)\, f.
	\end{equation}
	The limiting problem at $\varepsilon = \varepsilon^*$ reads
	\begin{equation}\label{eq:limiting_S0}
		L_1^0 f = \sigma\, P_\varphi f = \frac{\sigma\,\langle\varphi, f\rangle}{\|\varphi\|^2}\,\varphi.
	\end{equation}

	\emph{Step~3: Solution of the limiting problem.}
	Since $L_1^0$ is invertible on $\mathcal{S}_0$, we solve~\eqref{eq:limiting_S0} directly. Imposing $\langle\varphi, f\rangle = 1$:
	\begin{equation}
		f = \frac{\sigma}{\|\varphi\|^2}\,(L_1^0)^{-1}\varphi.
	\end{equation}
	The normalisation $\langle\varphi, f\rangle = 1$ determines
	\begin{equation}
		1 = \frac{\sigma}{\|\varphi\|^2}\,\langle\varphi,\,(L_1^0)^{-1}\varphi\rangle, \qquad \text{whence} \qquad \sigma_0^{(0)} = \frac{\|\varphi\|^2}{\langle\varphi,\,(L_1^0)^{-1}\varphi\rangle}.
	\end{equation}
	This is well-defined because $\langle\varphi,\,(L_1^0)^{-1}\varphi\rangle > 0$ by Lemma~\ref{condicion_no_degeneracion}.

	\emph{Step~4: Implicit function theorem.}
	Since $\ker(L_1^0) = \{0\}$, the map $$\Psi(\varepsilon, \sigma, f) := L_1^0\,f - \sigma\,P_\varphi f + (\varepsilon - \varepsilon^*)\sigma\,R_0(\varepsilon)\,f$$ restricted to the subspace $\{f \in \mathcal{S}_0 : \langle\varphi, f\rangle = 1\}$ satisfies $\Psi(\varepsilon^*, \sigma_0^{(0)}, f_0) = 0$ with $f_0 = \frac{\sigma_0^{(0)}}{\|\varphi\|^2}(L_1^0)^{-1}\varphi$. The tangent space to the constraint is $W_0 := \{v \in \mathcal{S}_0 : \langle\varphi, v\rangle = 0\}$. The joint Fr\'echet derivative $D_{(\sigma, f)}\Psi \colon \mathbb{R} \times W_0 \to \mathcal{S}_0$ at $(\varepsilon^*, \sigma_0^{(0)}, f_0)$ acts as
	\begin{equation}
		D_{(\sigma, f)}\Psi\cdot(\delta\sigma, \delta f) = L_1^0\,\delta f - \frac{\delta\sigma}{\|\varphi\|^2}\,\varphi.
	\end{equation}
	Given any $v \in \mathcal{S}_0$, we seek $(\delta\sigma, \delta f) \in \mathbb{R} \times W_0$ satisfying $L_1^0\,\delta f = v + \frac{\delta\sigma}{\|\varphi\|^2}\,\varphi$. Since $L_1^0$ is invertible, $\delta f = (L_1^0)^{-1}(v + \frac{\delta\sigma}{\|\varphi\|^2}\varphi)$, and the constraint $\langle\varphi, \delta f\rangle = 0$ determines
	\begin{equation}
		\delta\sigma = -\frac{\|\varphi\|^2\,\langle\varphi,(L_1^0)^{-1}v\rangle}{\langle\varphi,(L_1^0)^{-1}\varphi\rangle},
	\end{equation}
	which is well-defined since $\langle\varphi,(L_1^0)^{-1}\varphi\rangle > 0$. Hence $D_{(\sigma,f)}\Psi$ is a bijection, and the implicit function theorem \citep{kato:1995} guarantees the existence of unique smooth functions $\sigma(\varepsilon)$ and $f(\varepsilon)$ with $\sigma(\varepsilon^*) = \sigma_0^{(0)}$.

	\emph{Step~5: Conclusion.}
	Since $\Lambda^2 = (\varepsilon - \varepsilon^*)\,\sigma(\varepsilon)$ with $\sigma(\varepsilon^*) = \sigma_0^{(0)}>0$, then $\Lambda^2 < 0$ for $\varepsilon < \varepsilon^*$ (purely imaginary eigenvalues) and $\Lambda^2 > 0$ for $\varepsilon > \varepsilon^*$ (real eigenvalues, instability appears).
\end{proof}

We conclude this section by identifying explicit parameter ranges where the system exhibits instability, based on the Hamiltonian eigenvalue index theory \citep{kapitula:2013,pelinovsky:2005}.


\begin{proposition}[Instability intervals on $\mathcal{S}_0$]\label{prop:instability_intervals}
	Let $n \geq 1$ be such that $\lambda_{2n-1}^{(0)} < 1$. Then, for every
	\begin{equation}\label{eq:instability_interval}
		\varepsilon \in \big(\lambda_{2n-1}^{(0)},\; \min(1, \lambda_{2n}^{(0)})\big),
	\end{equation}
	the restricted system on $\mathcal{S}_0$ possesses at least one positive real eigenvalue $\Lambda > 0$. If $\lambda_{2n}^{(0)}$ does not exist, the interval in~\eqref{eq:instability_interval} is understood as $(\lambda_{2n-1}^{(0)}, 1)$.

\end{proposition}

\begin{proof}
	We apply the Hamiltonian eigenvalue count formula on $\mathcal{S}_0$, following \citet{pelinovsky:2005}:
	\begin{equation}\label{eq:count_formula}
		N_{\mathrm{real}} + 2N_{\mathrm{complex}} + 2N_{\mathrm{imag}}^- = n(L_0^0) + n(L_1^0),
	\end{equation}
	Here $n(\cdot)$ denotes the Morse index, that is, the number of negative eigenvalues counted with multiplicity. 
	Moreover, $N_{\mathrm{real}}$ counts positive real eigenvalues of the eigenvalue problem \eqref{eq:9kappa}-\eqref{eq:9kappaa}, $N_{\mathrm{complex}}$ counts complex eigenvalues in the first quadrant, 
	and $N_{\mathrm{imag}}^-$ counts purely imaginary eigenvalues with positive imaginary part and negative Krein signature with respect to $(L_0^0)^{-1}$.

	By Lemma~\ref{lem:L1_spectral}(i), we have $n(L_1^0)=1$. On the other hand, since $L_0^0 = \tilde{L}_0^{0} - \varepsilon$,
	the operator $L_0^0$ has a negative eigenvalue each time an eigenvalue $\lambda_k^{(0)}$ of $\tilde{L}_0^{0}$ satisfies $\lambda_k^{(0)} < \varepsilon$. Therefore,
	\[
		n(L_0^0) = \#\{k \geq 0 : \lambda_k^{(0)} < \varepsilon\}.
	\]

	Now let $\varepsilon \in (\lambda_{2n-1}^{(0)}, \min(1,\lambda_{2n}^{(0)}))$. Then
	\[
		\lambda_0^{(0)}, \lambda_1^{(0)}, \dots, \lambda_{2n-1}^{(0)} < \varepsilon,
	\]
	while either $\lambda_{2n}^{(0)} \geq \varepsilon$ or $\lambda_{2n}^{(0)}$ does not exist. Hence exactly $2n$ eigenvalues of $\tilde{L}_0^{0}$ lie below $\varepsilon$, and so $n(L_0^0)=2n$. Substituting this and $n(L_1^0)=1$ into~\eqref{eq:count_formula}, we obtain
	\[
		N_{\mathrm{real}} + 2N_{\mathrm{complex}} + 2N_{\mathrm{imag}}^- = 2n+1.
	\]
	The last two terms on the left-hand side are even integers, so they cannot by themselves produce the odd value $2n+1$. Consequently, $N_{\mathrm{real}}$ must be odd, in particular $N_{\mathrm{real}} \geq 1$. This yields a positive real eigenvalue $\Lambda > 0$.
\end{proof}

\begin{remark}\label{rem:instability_remark}
	For $\varepsilon \in (\lambda_{2n-2}^{(0)}, \lambda_{2n-1}^{(0)})$ ($n(L_0^0)$ even), the right-hand side of~\eqref{eq:count_formula} is even and does not force $n_{\mathrm{real}} \geq 1$; stability must be confirmed by direct spectral computation.
\end{remark}


\section{Numerical results}\label{sec:numerics}

The aim of this section is to conclude the theoretical analysis of the previous sections with numerical 
computations regarding the stability problem. To validate our analytical predictions, 
we discretise the radial spatial operators $L_0^l$ and $L_1^l$ with a sparse matrix approach based on the 
finite difference method, ensuring accurate resolution of both the discrete and the continuous spectrum. 
The numerical study is performed on the radial sector $\mathcal{S}_0$, the dipolar sector $\mathcal{S}_1$, 
and the full linearised operator, so that different perturbation classes and parameter regimes can be compared 
on the same footing. For this purpose, we compute the reduced ground state $\phi$ and the eigenvalues 
$\Lambda$ for dimensions $N=1,2,3$ and various values of $\kappa$. For every parameter set tested, the numerically computed spectrum of $\Psi_-$ contains a positive real 
eigenvalue, in agreement with the analytical prediction of Theorem~\ref{thm:psi_minus}.

\begin{figure}[h!]
	\centering
	\includegraphics[width=0.7\linewidth]{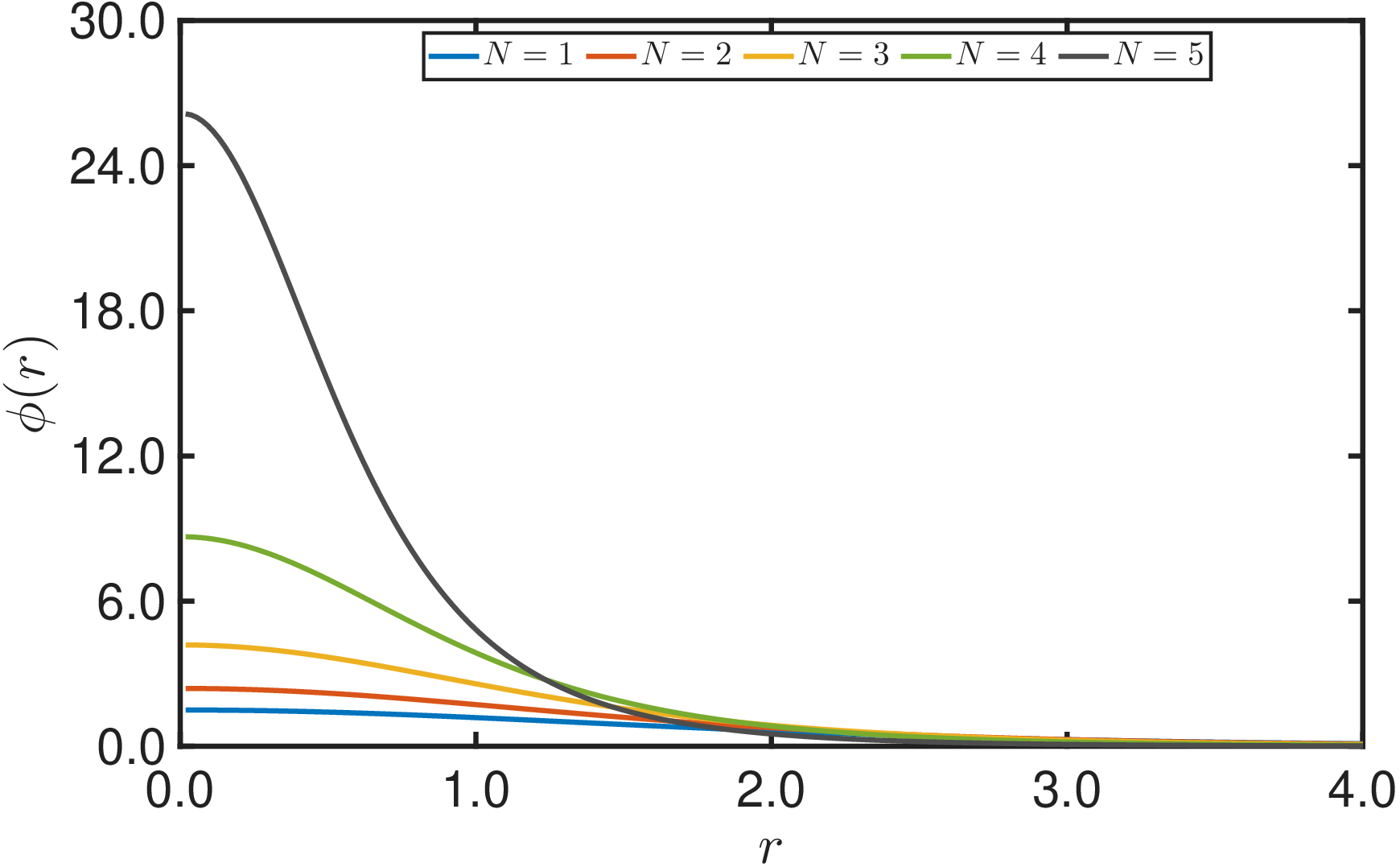}
	\caption{Ground states $(\phi)$ computed numerically for dimensions $N = 1, 2, 3, 4, 5$ and $\kappa = 0.5$. Parameters: $R = 50$, $N_x = 800$}
	\label{fig:solitonesk05}
\end{figure}

Since $\phi(r)$ is a radial positive solution of~\eqref{eq:psi_rescaled}, it satisfies the ODE
\begin{equation}\label{ecuacion_radial}
	-\frac{\partial^2\phi}{\partial r^2} - \frac{N-1}{r}\frac{\partial\phi}{\partial r} + \phi - \phi^{2\kappa+1} = 0, \qquad r > 0.
\end{equation}
Regularity at the origin requires $\phi'(0) = 0$. We discretise $(0,R]$ with step size $h$ and approximate the derivatives by centred finite differences:
\begin{equation}
	D_2\phi(r) = \frac{\phi(r+h) - 2\phi(r) + \phi(r-h)}{h^2}, \qquad D_1\phi(r) = \frac{\phi(r+h) - \phi(r-h)}{2h}.
\end{equation}
The nonlinear system is solved by a Newton--Raphson iteration with Jacobian freezing; see, for example, \citep{press:1992,ortega:1970}. Unlike the one-dimensional case, no closed-form solution is available as initial seed. For this purpose, we use the ansatz
\begin{equation}
	\phi_0(r) = A\,\operatorname{sech}^{1/\kappa}(B\kappa r)\frac{1}{1 + (Br)^{(N-1)/2}},
\end{equation}
where $A$ and $B$ are determined by the Pohozaev identities~\eqref{poho1}--\eqref{poho2} (after the variable change). This seed yields convergence for all dimensions tested ($N = 1, \ldots, 10$) with residual errors of order~$10^{-12}$ (see Fig.~\ref{fig:solitonesk05} where $\phi$ is plotted).

\subsection{Sufficient condition for instability and stability diagrams}

We present a sufficient condition for instability arising from bifurcation at $\Lambda = 0$. We compute the discrete eigenvalues of $\tilde{L}_0^{0}$ lying in $[0,1]$ and the critical value $\varepsilon^\mu$ defined in Theorem~\ref{thm:direction_splitting_S1} as the first value of $\varepsilon$ for which $\mu(\varepsilon) = 0$.

\begin{figure}
	\centering
	\includegraphics[width=0.51\linewidth]{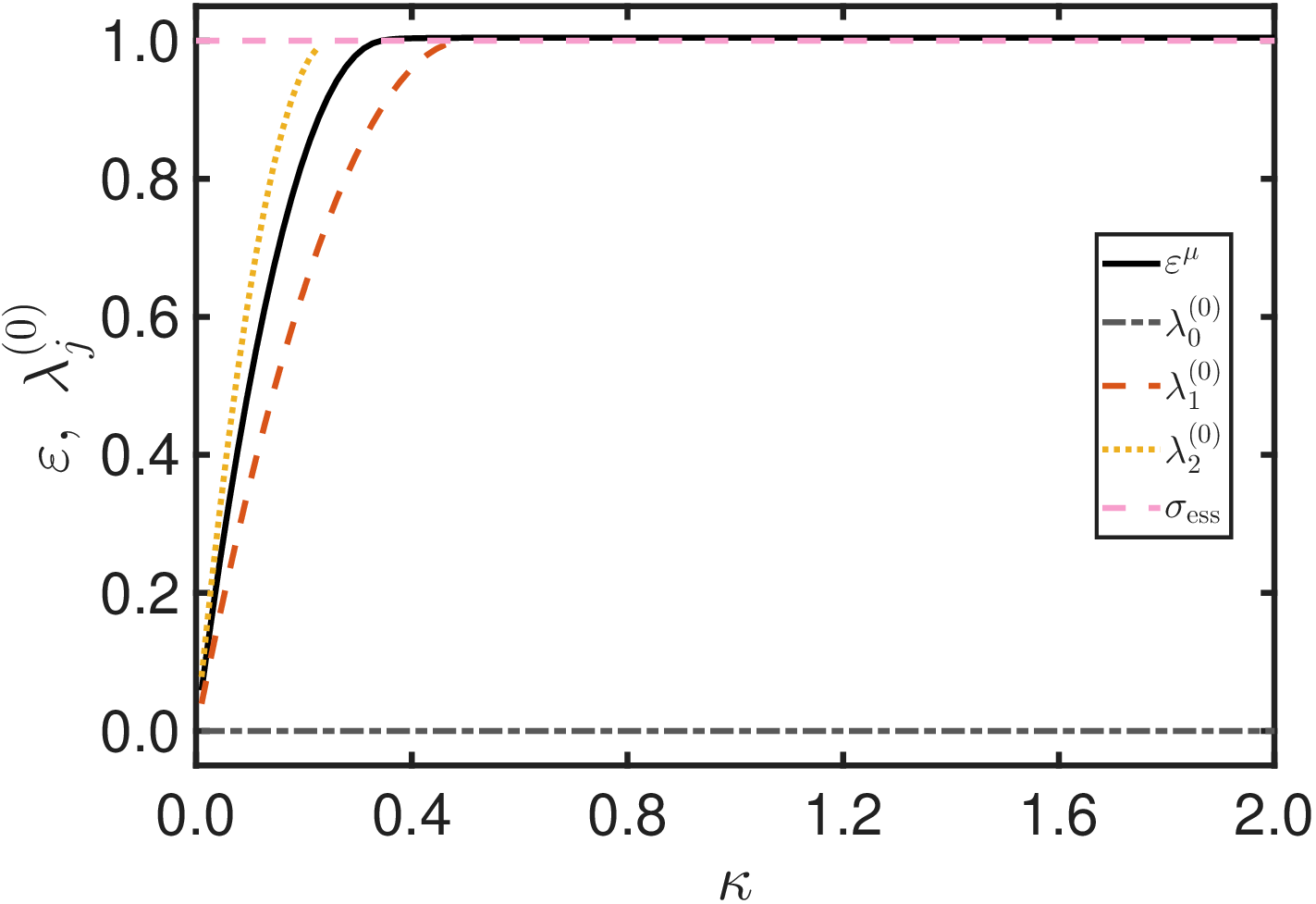}
	\includegraphics[width=0.51\linewidth]{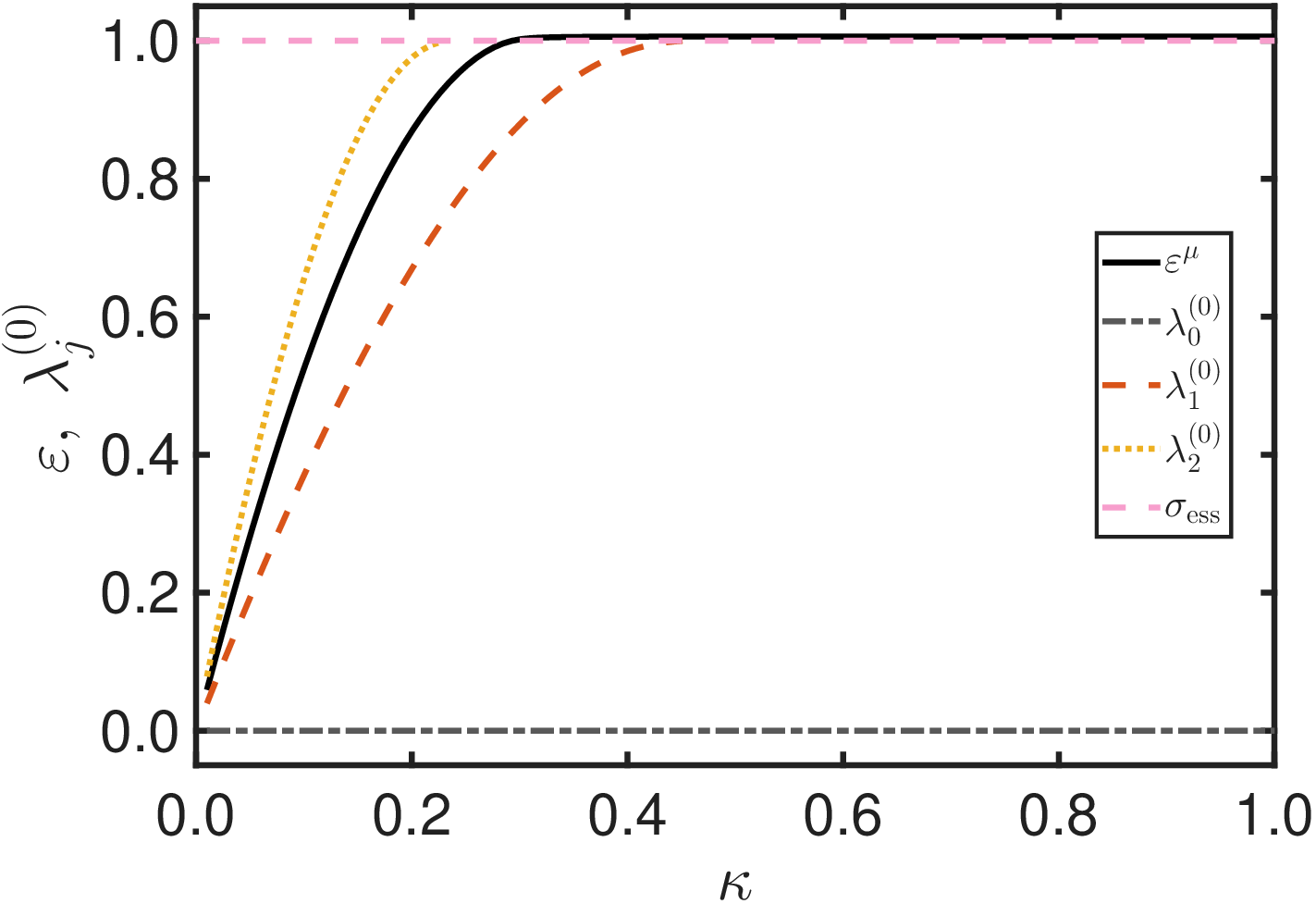}
	\includegraphics[width=0.51\linewidth]{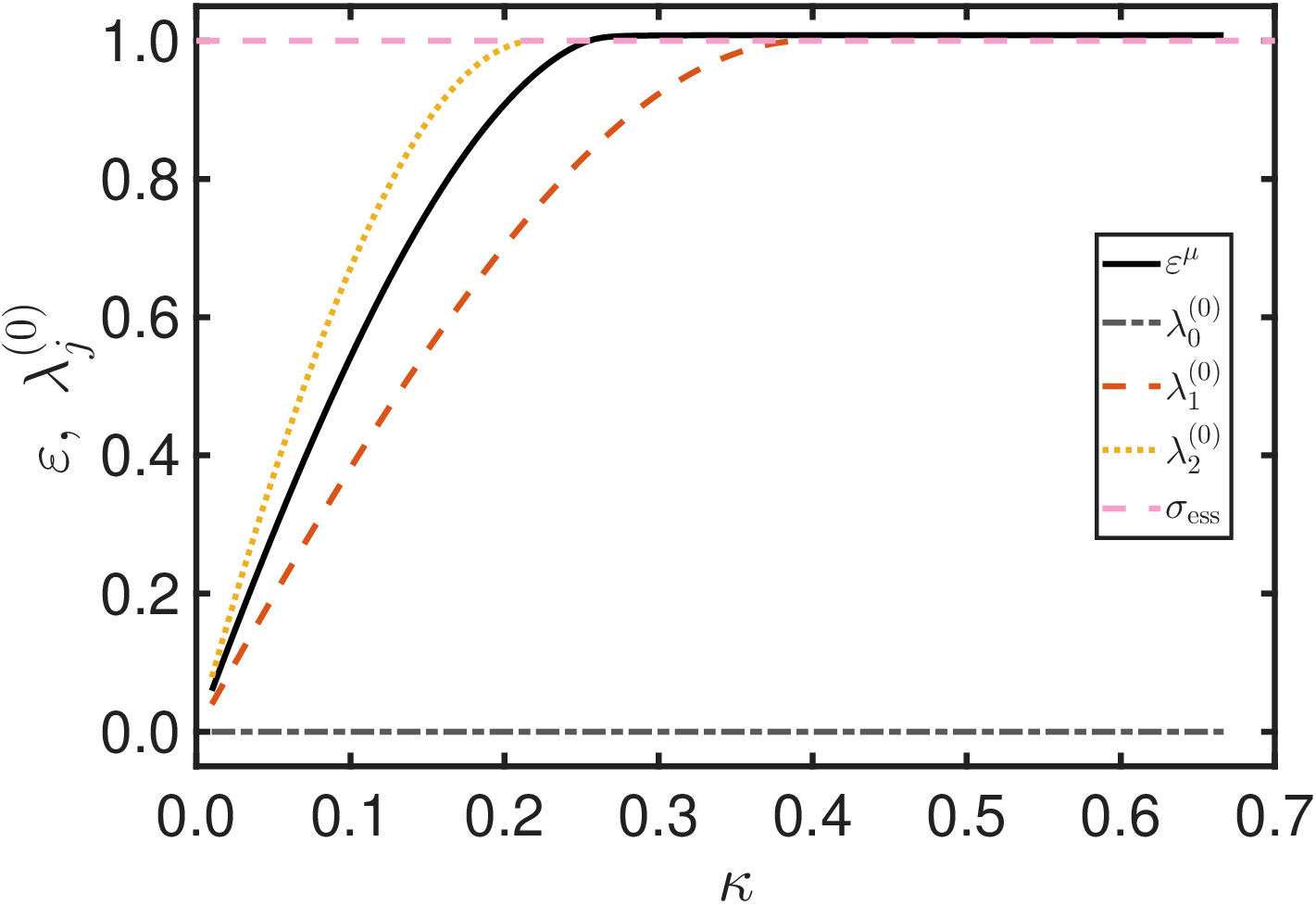}
	\caption{Discrete eigenvalues $\lambda_n^{(0)}$ of $\tilde{L}_0^{0}$ in the interval $[0.01,2/N]$ (dashed and dotted curves), together with the critical value $\varepsilon^\mu$ defined by $\mu(\varepsilon^\mu)=0$ (solid black curve), plotted as functions of~$\kappa$ for $N = 1$ (first), $N = 2$ (second), and $N = 3$ (third). As $\kappa$ varies, the eigenvalue branches enter and leave the interval $(0,1)$. In all cases, the ordering $\lambda_1^{(0)} < \varepsilon^\mu < \lambda_2^{(0)}$ holds throughout the plotted range, showing that the $\mathcal{S}_0$-bifurcation precedes the $\mathcal{S}_1$-bifurcation}
	\label{fig:eigenvalues_L0}
\end{figure}

According to the numerical computation for dimensions $N = 1, 2, 3$ (see Fig.~\ref{fig:eigenvalues_L0}), we obtain the ordering $\lambda_1^{(0)} < \varepsilon^\mu < \lambda_2^{(0)}$ for all values of $\kappa$ and $N$ tested. This ordering has a decisive consequence: combining Corollary~\ref{corolario_bifurcacion}, Theorem~\ref{thm:bifurcation_S0}, and Proposition~\ref{prop:instability_intervals}, we conclude that the $\mathcal{S}_0$-bifurcation at $\varepsilon = \lambda_1^{(0)}$ always produces instability \emph{before} the $\mathcal{S}_1$-bifurcation at $\varepsilon = \varepsilon^\mu$. In other words, the radial sector $\mathcal{S}_0$ governs the onset of instability: for $\varepsilon > \lambda_1^{(0)}$, the system is unstable because the spectrum contains a real eigenvalue with positive real part.

We also find numerically that for $\kappa \geq \kappa_{\mathrm{crit}} = 2/N$, there is no eigenvalue of $\tilde{L}_0^{0}$ in the interval $(0,1)$, so no bifurcation at $\Lambda = 0$ on $\mathcal{S}_0$ occurs in the admissible range. In this regime, the stability analysis reduces to the numerical determination of the complex quadruplet and the condition $|\Lambda_i| > |\Lambda_r|$ from~\eqref{neccesary_condition_stability}.

For values of $\varepsilon \leq \lambda_1^{(0)}$, we need to numerically compute the spectrum of $\mathcal{L}$ to determine stability. According to the count formula~\eqref{eq:count_formula}, for $\varepsilon < \lambda_1^{(0)}$ (where only the zero eigenvalue $\lambda_0^{(0)}=0$ lies below $\varepsilon$), {the Morse index is} $n(L_0^0) = 1$ and $n(L_1^0) = 1$, giving a total of~$2$. This implies that instability can only arise from a pair of purely real positive eigenvalues or from a single complex quadruplet, which simplifies the numerical search.

In dimension $N = 1$, the eigenvalues of $\tilde{L}_0$ can be analytically calculated (see Appendix~\ref{app:eigenvalues_dim1}):
\begin{equation}\label{eq:eigenvalues_L0_tilde_1d}
	\lambda_n = \kappa n(2 - \kappa n), \qquad \kappa <\frac{1}{n},
\end{equation}
and the first excited state of even parity (corresponding to the subspace $\mathcal{S}_0$) has eigenvalue $\lambda_1^{(0)} = \lambda_2 = 4\kappa(1-\kappa)$ with $\kappa < 1/2$. For $\kappa \geq 1/2$, no eigenvalue of $\tilde{L}_0^{0}$ lies in $(0,1)$, so no bifurcation at zero occurs on $\mathcal{S}_0$, which corresponds to the one-dimensional case described by \citet{carreno:2024}.

The existence of a critical value $\varepsilon = \lambda_1^{(0)}$ defines a new stability boundary in the $(\rho, r)$--plane which is given by the curve
\begin{equation}
	r = \sqrt{\rho^2 + \left(\frac{\lambda_1^{(0)}}{2 - \lambda_1^{(0)}}\right)^2},
\end{equation}
which can be easily obtained using Eq.~\eqref{eq:stability_curve}.

The region where stability might appear is therefore restricted to $\varepsilon \leq \lambda_1^{(0)}$ if it exists and $\varepsilon < 1$ otherwise. In this region, we distinguish two regimes:

	\begin{enumerate}[label=(\roman*)]
		\item \textbf{Quadruplet-free baseline and low-$\kappa$ stable windows:} 
		\begin{itemize}
			\item For $\varepsilon \leq \varepsilon_{\text{quad}}$, no complex quadruplet appears in the spectrum of $\mathcal{L}$, and stability is determined solely by whether $\kappa < 2/N$ (stable) or $\kappa \geq 2/N$ (unstable).
			\item For small values of $\kappa < 2/N$, an additional stable region emerges enclosed between the curves $\lambda_1^{(0)}(\kappa)$ and $\varepsilon_{\text{spike}}(\kappa)$, where the system displays stability without requiring the damping mechanism (see Fig.~\ref{fig:fig1}). This region accounts for the stability transition windows observed in the $(\rho, r)$ phase planes for low nonlinearity.
		\end{itemize}
		
		\item \textbf{Damping-controlled regime:} For $\varepsilon > \varepsilon_{\text{quad}}$ and outside the region enclosed between the curves $\lambda_1^{(0)}(\kappa)$ and $\varepsilon_{\text{spike}}(\kappa)$, a complex quadruplet appears in the spectrum of $\mathcal{L}$, and stability depends on the relative sizes of the real and imaginary parts of the eigenvalues (cf.~\eqref{neccesary_condition_stability}):
		\begin{itemize}
			\item For $\varepsilon_{\text{quad}} < \varepsilon \leq \varepsilon_c$, the damping is insufficient to compensate for the oscillatory instability, and the system remains unstable.
			\item For $\varepsilon > \varepsilon_c$, the damping is sufficient to stabilise the system (see Fig.~\ref{fig:fig1}). In particular, for $\kappa < 2/N$, one has $\varepsilon_c = \varepsilon_{\text{quad}}$ (see Appendix~\ref{app:subcritical_quadruplets}), so the system can be stabilised as soon as $\varepsilon > \varepsilon_{\text{quad}}$.
		\end{itemize}
	\end{enumerate}

\begin{figure}[h!]
    \centering
    \includegraphics[width=0.49\linewidth]{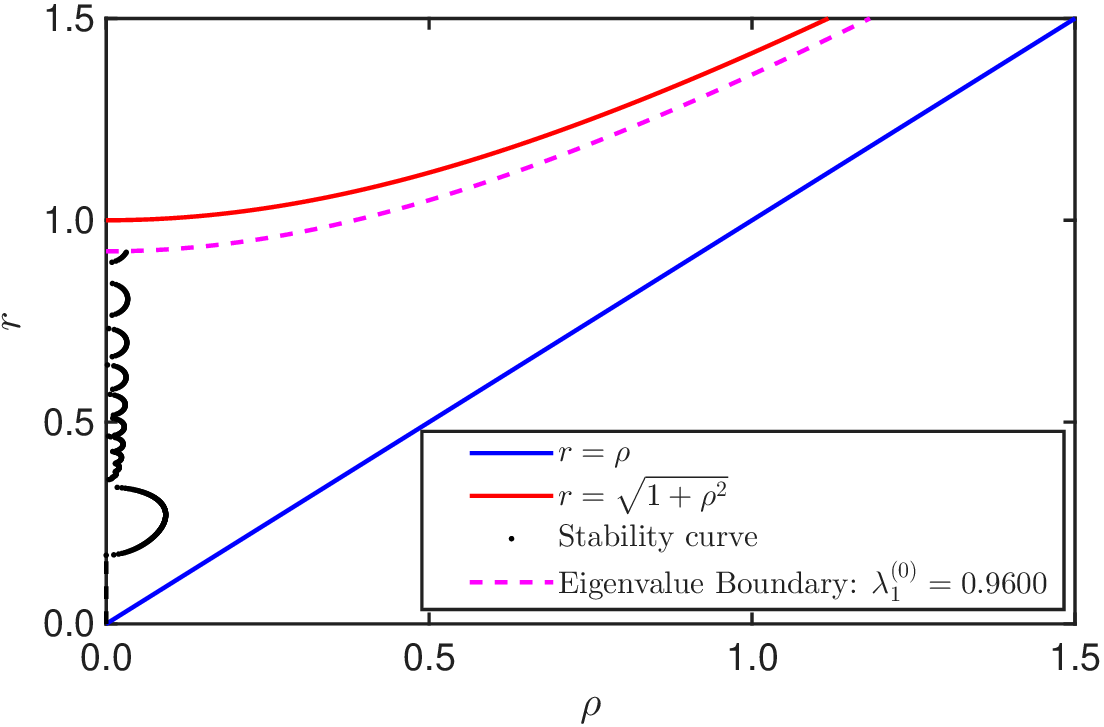}\hfill
    \includegraphics[width=0.49\linewidth]{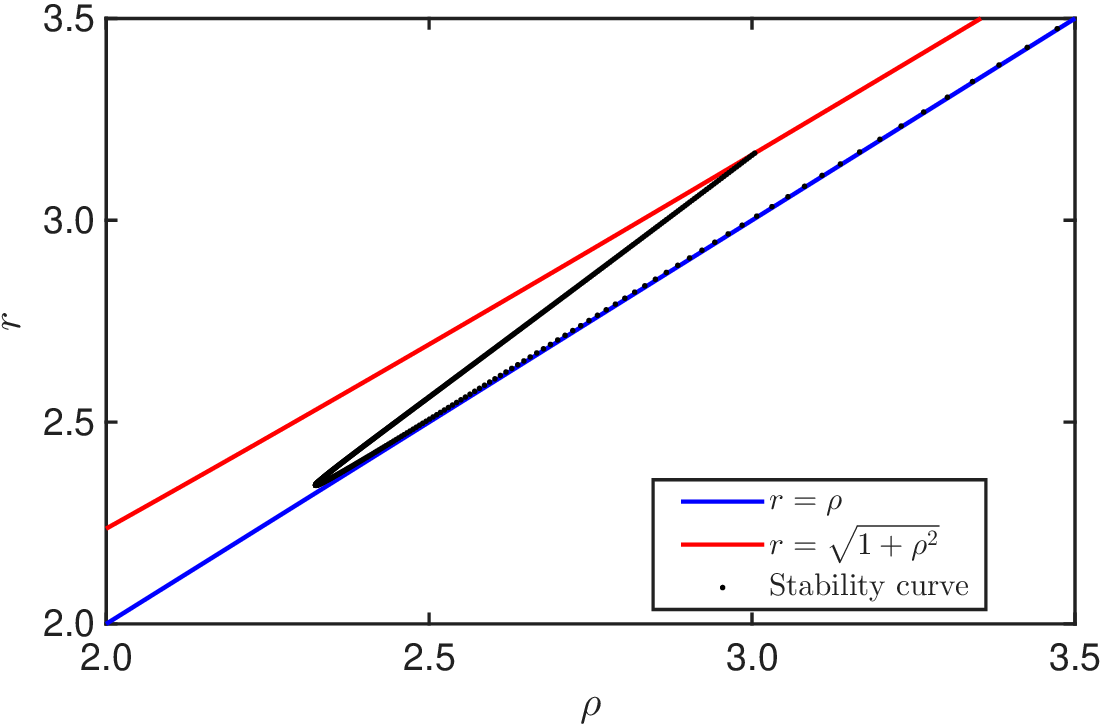}
    \vskip 3pt
    \includegraphics[width=0.49\linewidth]{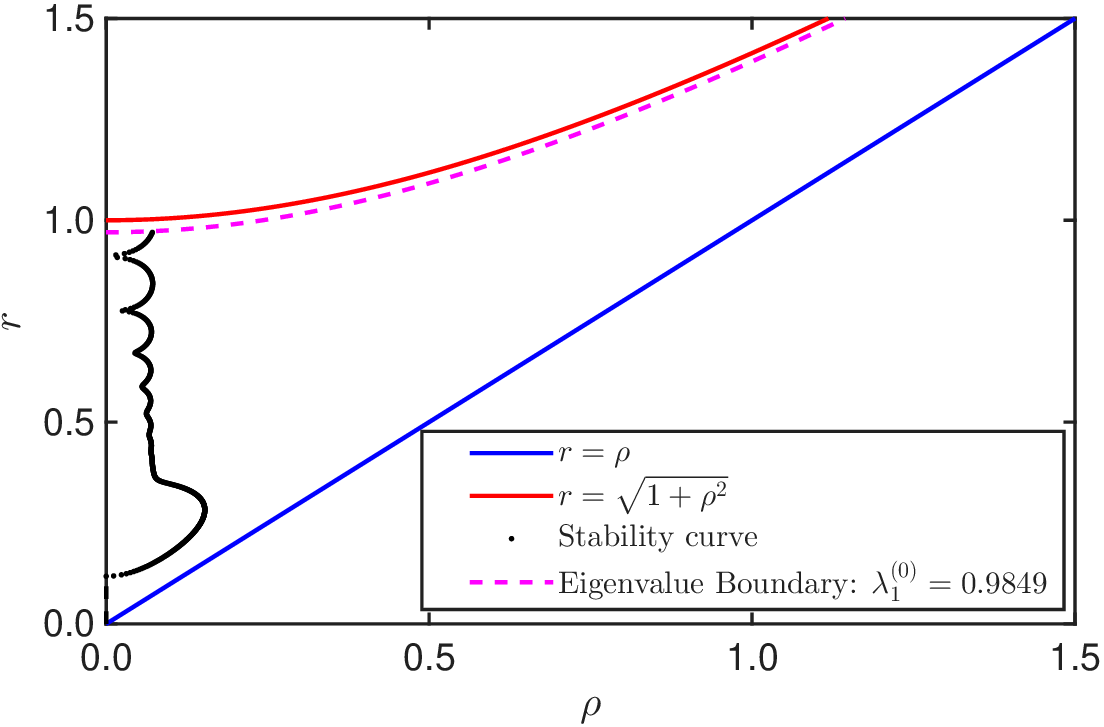}\hfill
    \includegraphics[width=0.49\linewidth]{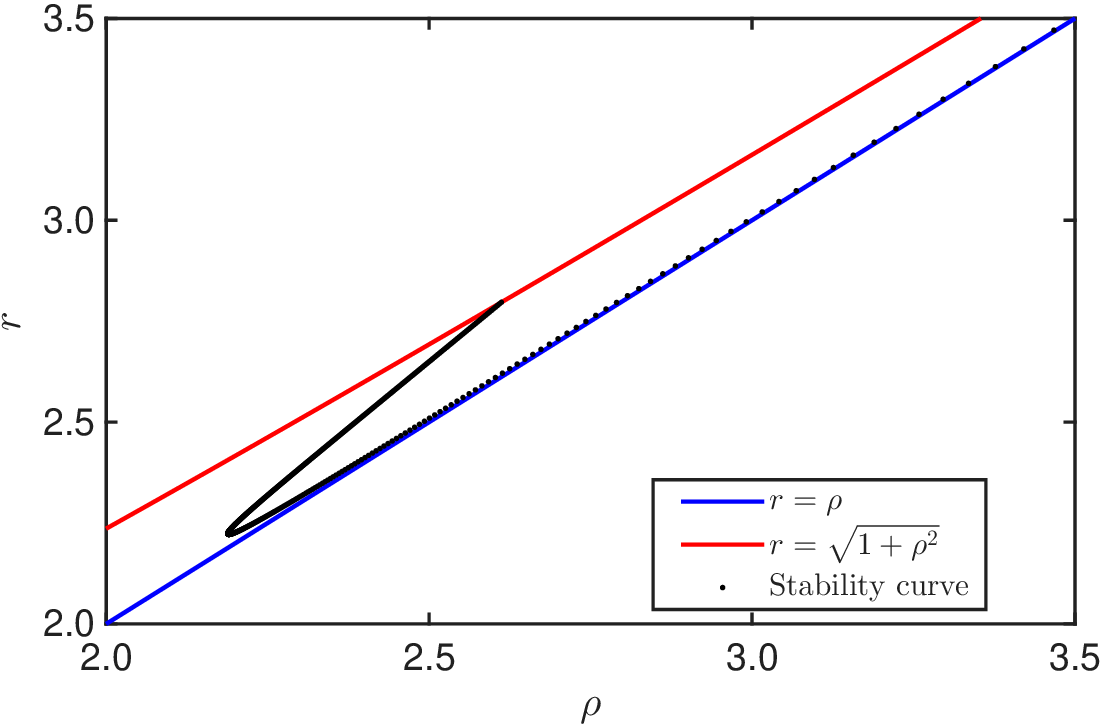}
    \vskip 3pt
    \includegraphics[width=0.49\linewidth]{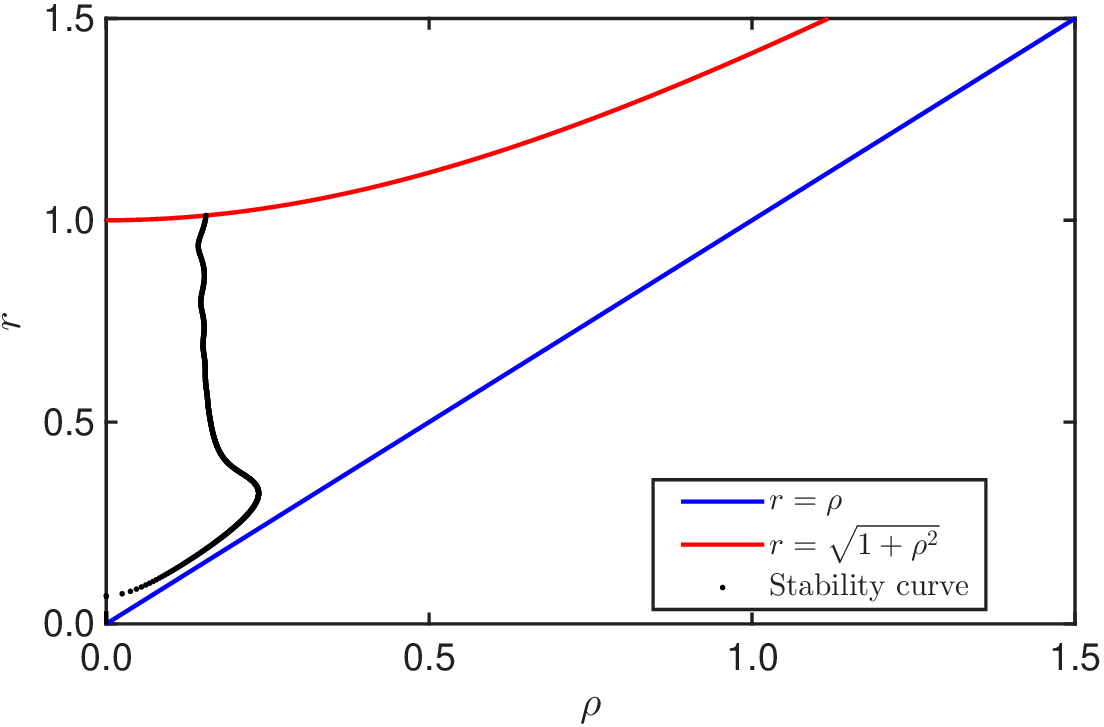}\hfill
    \includegraphics[width=0.49\linewidth]{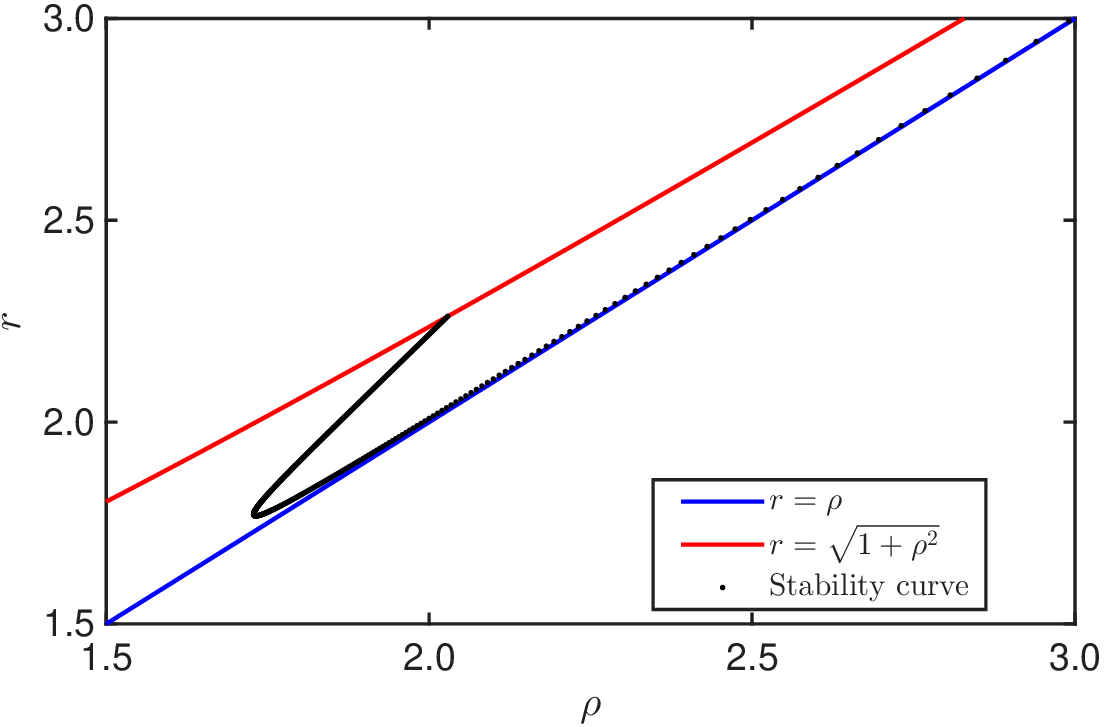}
    
    \caption{Stability diagrams in the $(\rho, r)$--plane for dimensions $N = 1$ (top), $N = 2$ (middle), and $N = 3$ (bottom). The left column corresponds to subcritical nonlinearity ($\kappa = 0.4$), whereas the right column features supercritical regimes satisfying $\kappa > 2/N$: specifically, $\kappa = 2.50$ ($N = 1$), $\kappa = 1.20$ ($N = 2$), and $\kappa = 0.75$ ($N = 3$). The blue solid line represents the existence boundary $r = \rho$, below which no soliton solutions exist. The red solid line marks the essential spectrum boundary $r = \sqrt{1+\rho^2}$ (cf.~Section~\ref{sec:stability}). The black dotted curve denotes the stability boundary~\eqref{eq:stability_curve}; to its left, damping is insufficient to offset the oscillatory instability. The magenta dashed line indicates the upper theoretical boundary $\tilde{r} = \sqrt{\rho^2 + (\lambda_1^{(0)}/(2 - \lambda_1^{(0)}))^2}$ arising from the first excited eigenvalue of $\tilde{L}_0^{0}$; exceeding this curve induces instability via $\mathcal{S}_0$ bifurcation (Proposition~\ref{prop:instability_intervals}). The overall stable domain is delimited by the black dotted curve, the blue solid boundary, and the lower of the magenta dashed and red solid curves}
    \label{fig:n1k04rrho}
\end{figure}

Using the numerical procedure described at the beginning of the section, we obtain stability diagrams in the $(\rho, r)$--plane for $\kappa = 0.4$ and dimensions $N = 1, 2, 3$, shown in Fig.~\ref{fig:n1k04rrho}. Each diagram displays up to four curves that delimit the stability region:
\begin{enumerate}[label=(\roman*)]
	\item The \emph{existence boundary} $r = \rho$ ($\varepsilon = 0$), below which the soliton $\Psi_+$ does not exist.
	\item The \emph{essential spectrum boundary} $r = \sqrt{1 + \rho^2}$ ($\varepsilon = 1$), above which the essential spectrum of $\mathcal{L}$ reaches the origin and the soliton is unstable independently of the damping (Section~\ref{sec:stability}).
	\item The \emph{stability curve}~\eqref{eq:stability_curve}, parametrised by $\varepsilon \in [0,1]$ or up to the value $\lambda_1^{(0)}$.

	\item When $\kappa < \kappa_{\mathrm{crit}} = 2/N$, the \emph{upper eigenvalue boundary} $r = \sqrt{\rho^2 + (\lambda_1^{(0)}/(2 - \lambda_1^{(0)}))^2}$, above which the $\mathcal{S}_0$-bifurcation at $\varepsilon = \lambda_1^{(0)}$ renders the soliton unstable regardless of the damping (Proposition~\ref{prop:instability_intervals}).
\end{enumerate}
The \emph{stable region} is the area enclosed between the stability curve, the more restrictive of boundaries~(ii) and~(iv) and the existence region~(i).

The qualitative behaviour is similar across dimensions, with the critical nonlinearity exponent $\kappa_{\mathrm{crit}} = 2/N$ decreasing with~$N$ and the ground state ceasing to exist for $\kappa \geq 2/(N-2)$ when $N \geq 3$.

Another noteworthy feature is that the curve $\varepsilon_c(\kappa)$ increases throughout the supercritical regime $\kappa \geq 2/N$. Consequently, the damping required to stabilise the soliton becomes stronger as $\kappa$ increases, which is consistent with the fact that the nonlinearity becomes more focusing and hence more prone to instability.

In fact, we find a maximum value of $\kappa$, denoted by $\kappa_{\max}^N$, beyond which the stability zone vanishes entirely and no choice of $(\rho, r)$ within the existence region can stabilise the soliton. For the dimensions tested, we obtain $\kappa_{\max}^1 \approx 4.00$, $\kappa_{\max}^2 \approx 1.58$, and $\kappa_{\max}^3 \approx 0.93$. In particular, since $\kappa_{\max}^3 < 1$, the soliton of the cubic PDDNLS equation in dimension $N = 3$ cannot be stabilised and is therefore always unstable. This phenomenon is illustrated in the $(\kappa, \varepsilon)$--diagrams of Fig.~\ref{fig:fig1}.

\begin{figure}[h!]
    \centering
    \includegraphics[width=0.48\linewidth]{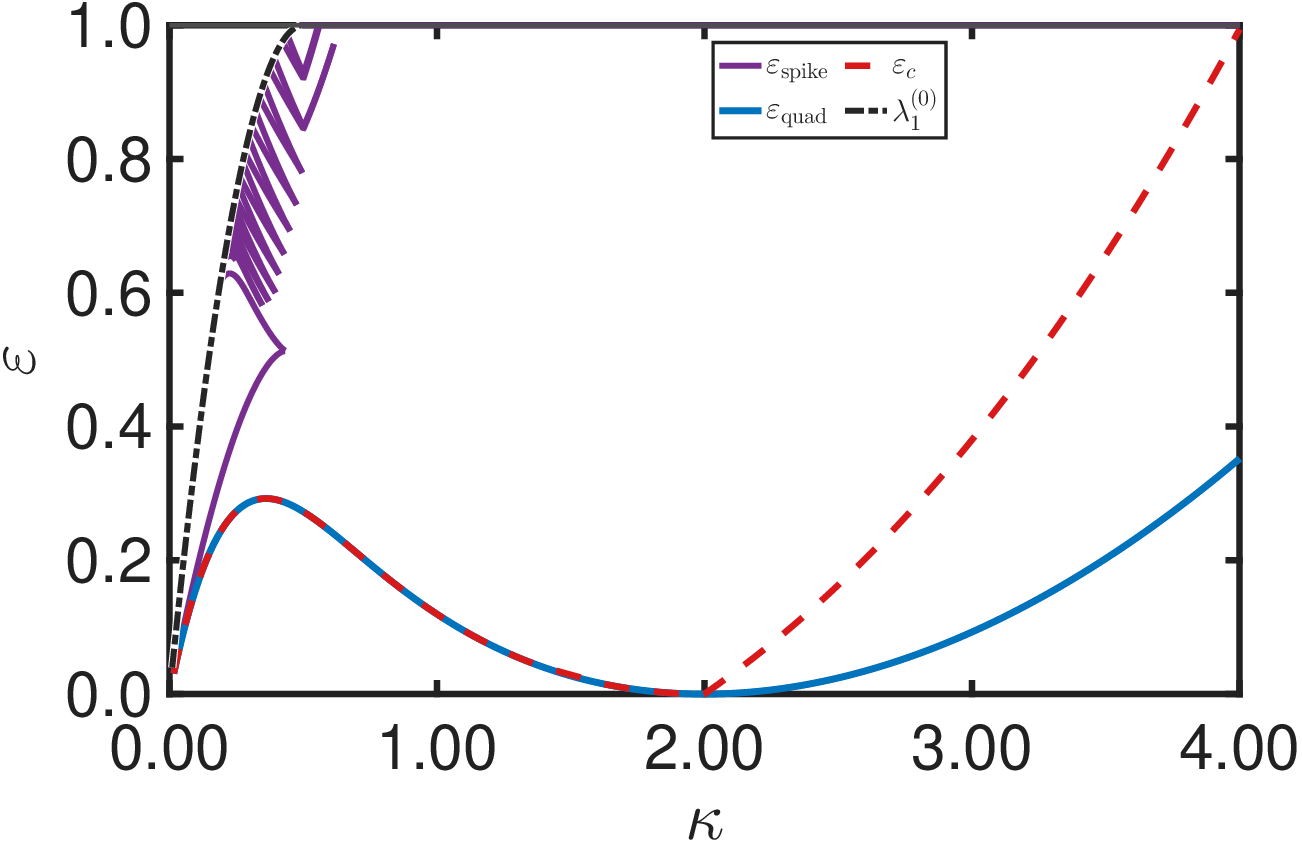}\hfill
    \includegraphics[width=0.48\linewidth]{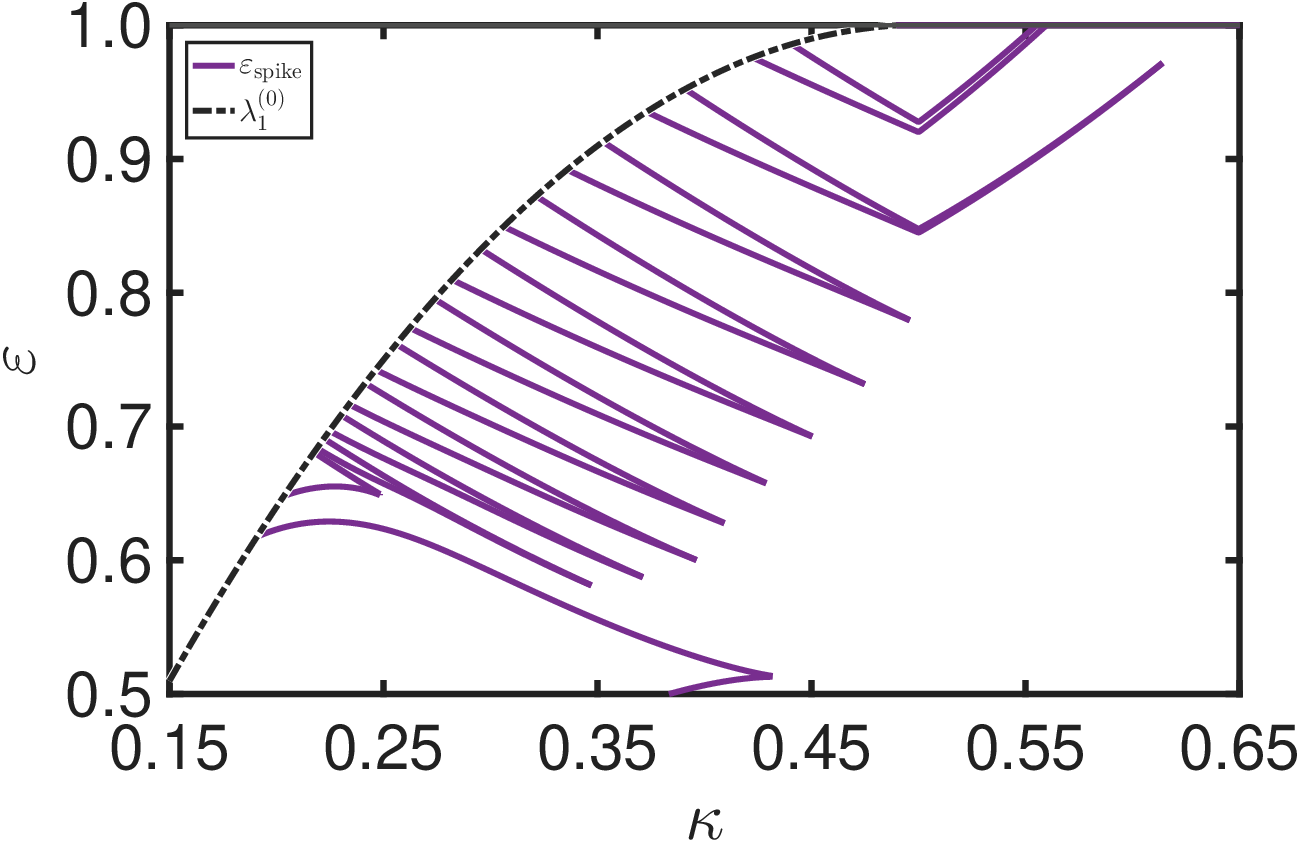}
    \vskip 2pt
    \includegraphics[width=0.48\linewidth]{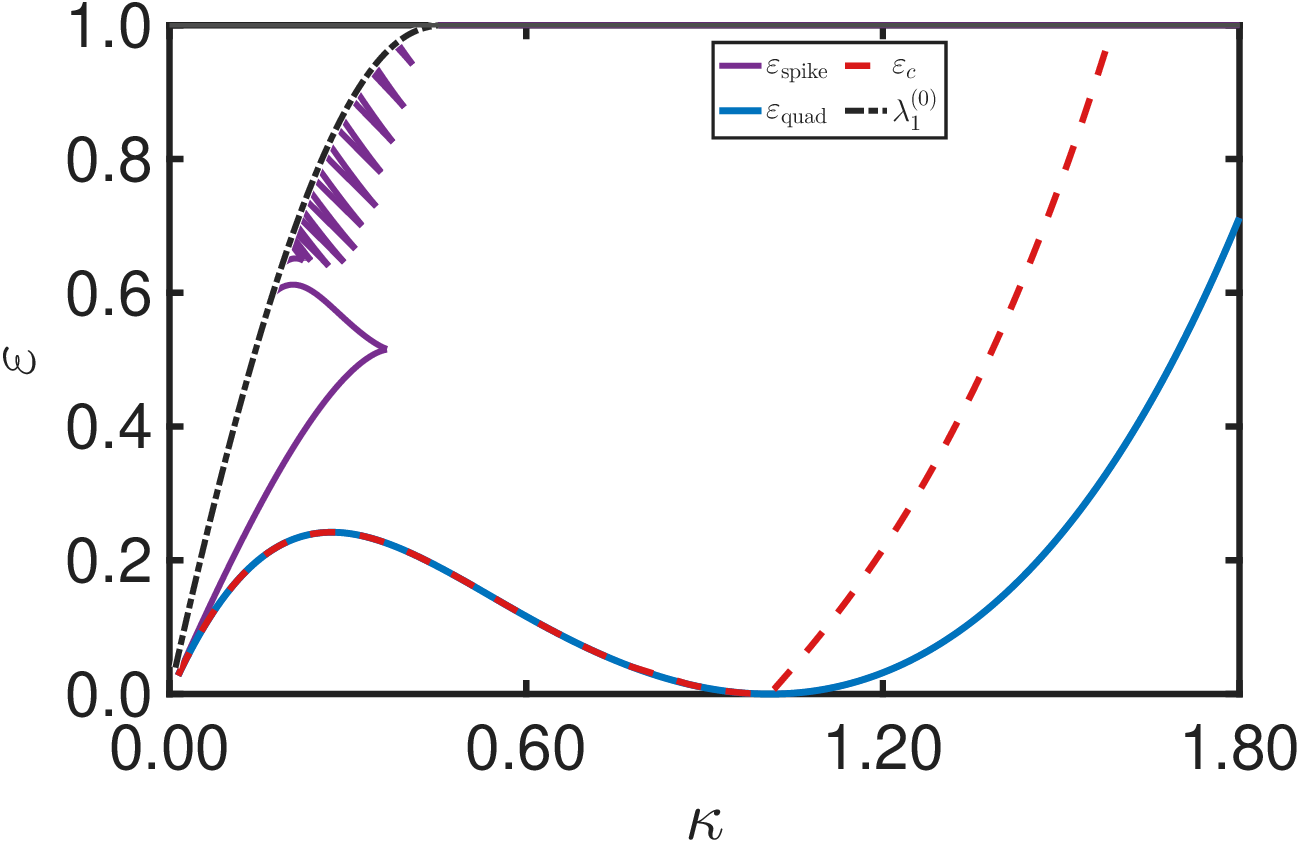}\hfill
    \includegraphics[width=0.48\linewidth]{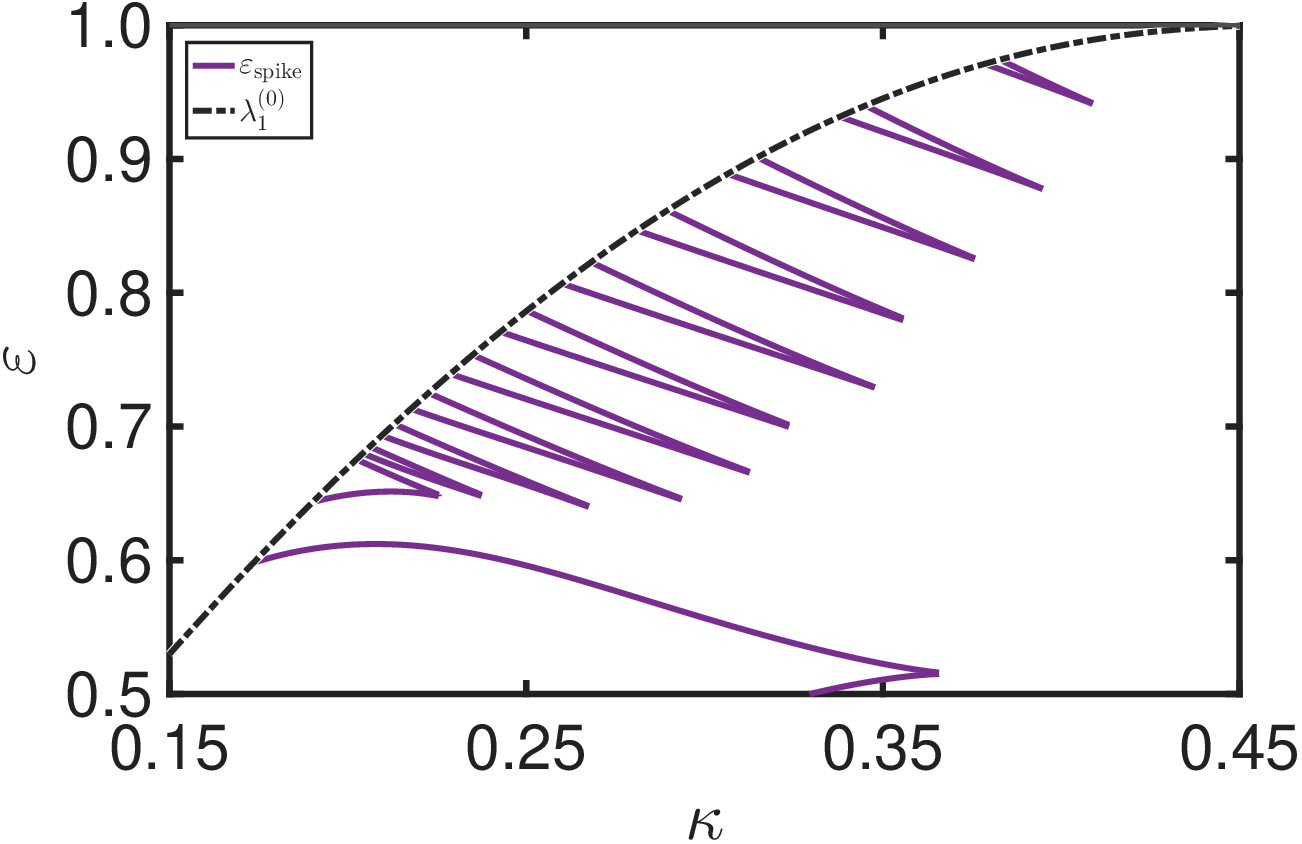}
    \vskip 2pt
    \includegraphics[width=0.48\linewidth]{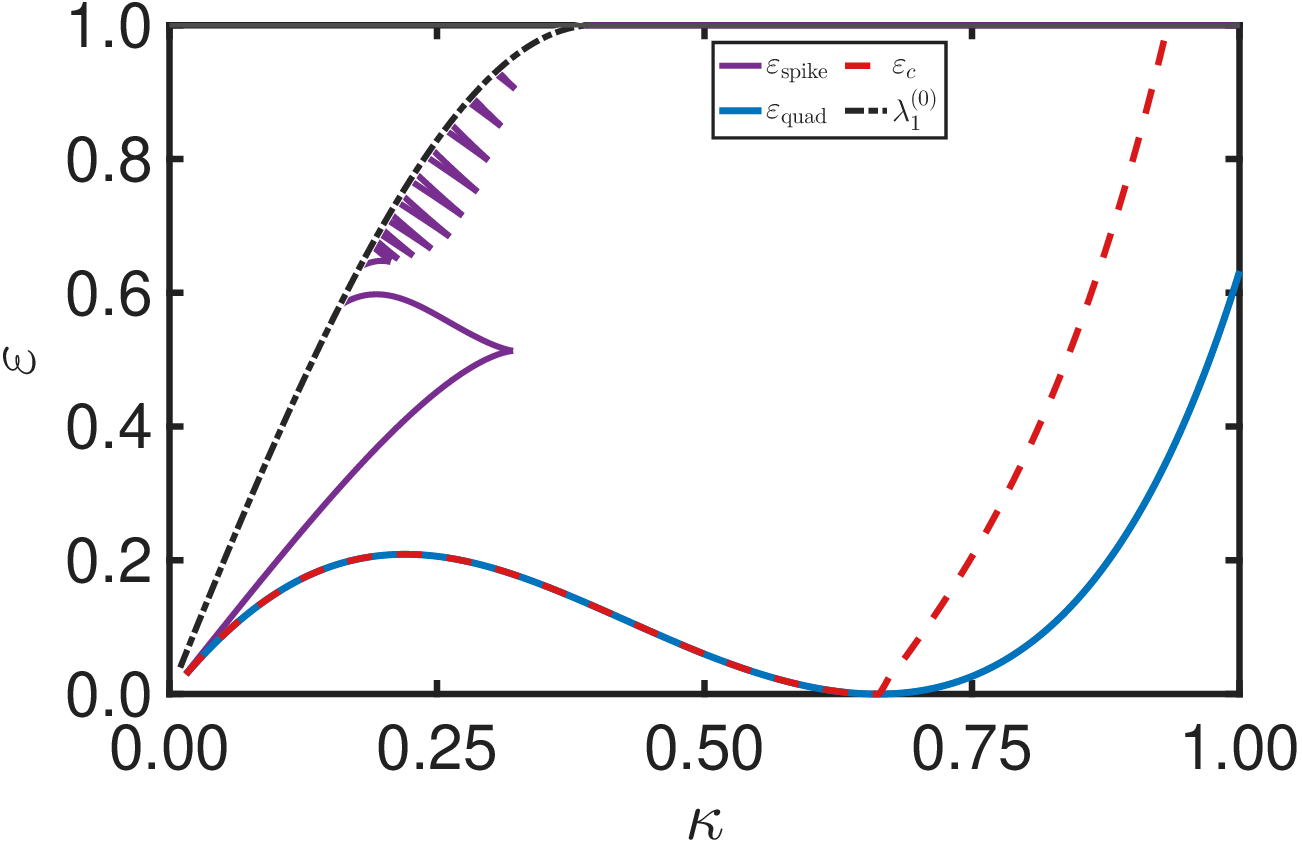}\hfill
    \includegraphics[width=0.48\linewidth]{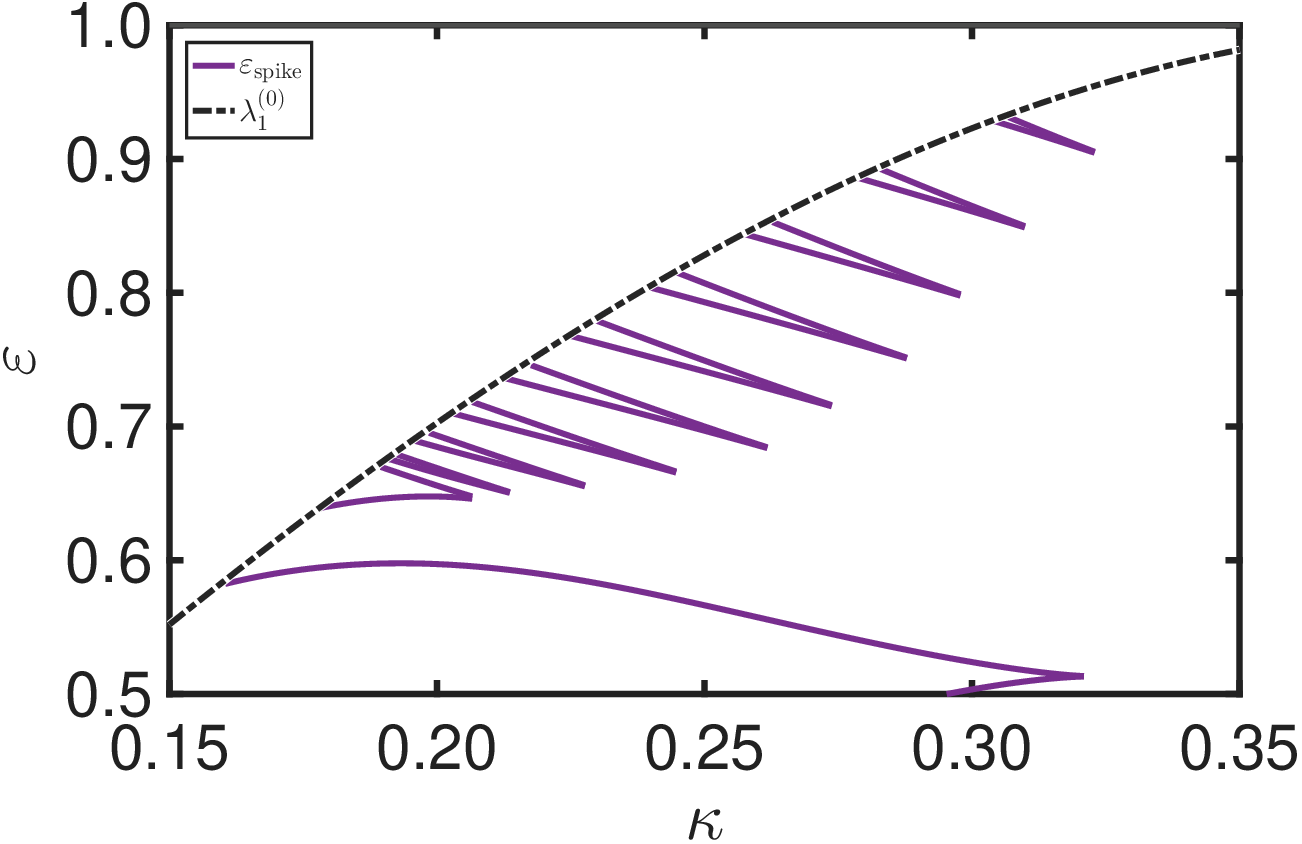}
    
\caption{$(\kappa,\varepsilon)$--stability diagrams for $N = 1$ (top), $N = 2$ (middle), and $N = 3$ (bottom), showing full parameter ranges (left) and zoomed-in views (right). 
Curves represent the onset of a complex quadruplet $\varepsilon_{\text{quad}}$ (solid blue), the lower boundary of stable windows $\varepsilon_{\text{spike}}$ (solid purple), the threshold $\varepsilon_c$ where $|\Lambda_r| = |\Lambda_i|$ (dashed red; necessary for stabilization via~\eqref{neccesary_condition_stability}), and the first excited eigenvalue $\lambda_1^{(0)}(\kappa)$ of $\tilde{L}_0^{0}$ (dash-dotted black), above which the system is unconditionally unstable (Proposition~\ref{prop:instability_intervals}). 
The \textit{stabilization region} comprises the parameter space where the system can be stabilized for a suitable choice of the dissipation parameter; this domain is delimited by $\varepsilon_c$ and $\varepsilon < 1$ in the supercritical regime ($\kappa > 2/N$), intersecting at $\kappa = \kappa_{\max}^N$ (beyond which stabilization is impossible), and by $\lambda_1^{(0)}$ and $\varepsilon < 1$ in the subcritical regime ($\kappa < 2/N$). 
In contrast, \textit{stable regions} correspond to parameter values where the system is stable independently of the dissipation; these are given by the narrow windows enclosed by $\varepsilon_{\text{spike}}$ and, in the subcritical regime, the region lying below $\varepsilon_{\text{quad}}$}
\label{fig:fig1}
\end{figure}

\section{Conclusions}\label{sec:summary}

We have extended the stability analysis of the parametrically driven and damped nonlinear Schr\"odinger equation from one to $N$ spatial dimensions. Two families of stationary solutions $\Psi_\pm$ have been identified; their existence conditions depend on the parameters $r$, $\rho$, $\kappa$ and $N$.

By linearising around $\Psi_\pm$ and introducing the parametrisation through~$\varepsilon$, the stability problem reduces to a dissipationless eigenvalue problem involving the operators $L_0$ and $L_1$. We proved that $\Psi_-$ is always unstable and that the stability of $\Psi_+$ is governed by the spectrum of these operators restricted to the angular momentum sectors $\mathcal{S}_l$.

The angular momentum decomposition reveals that:
\begin{enumerate}[label=(\roman*)]
	\item Modes with $l \geq 2$ never produce instabilities (Theorem~\ref{thm:stability_l_geq_2}).
	\item On $\mathcal{S}_1$, eigenvalues are real or purely imaginary, so no complex quadruplets arise (Proposition~\ref{prop:l1_structure}). The non-splitting criterion (Theorem~\ref{thm:no_splitting}) and the analysis of the two lowest eigenvalues of $\tilde{L}_0^{1}$ (Propositions~\ref{prop:no_splitting_S1} and~\ref{prop:no_splitting_S1_excited}) show that there is a unique critical value $\varepsilon^\mu$ at which the instability sets in, producing real eigenvalues for $\varepsilon > \varepsilon^\mu$ (Theorems~\ref{thm:direction_splitting_S1} and~\ref{thm:bifurcation_L0_eigenvalue}; degenerate case in Theorem~\ref{thm:degenerate_S1}).
	\item On $\mathcal{S}_0$, splitting from $\Lambda = 0$ occurs at every eigenvalue of $\tilde{L}_0^{0}$, producing instabilities as $\varepsilon$ increases from these points (Theorem~\ref{thm:bifurcation_S0}). By a numerical study of the eigenvalues of $\tilde{L}_0^{0}$ and $\varepsilon^\mu$, we conclude that the radial sector governs the full stability picture, simplifying the numerical study.
\end{enumerate}

Three new phenomena, absent from the one-dimensional analysis of \citet{carreno:2024}, have been identified:
\begin{itemize}
    \item \emph{Small $\kappa$ ($\kappa < \kappa_{\mathrm{crit}} = 2/N$):} Eigenvalues of $\tilde{L}_0^{0}$ enter the admissible range $(0,1)$ and trigger an additional instability via bifurcation at $\Lambda = 0$ on $\mathcal{S}_0$. This produces an upper stability boundary $r = \sqrt{\rho^2 + (\lambda_1^{(0)}/(2-\lambda_1^{(0)}))^2}$ in the $(\rho, r)$--plane, above which the soliton is unstable regardless of the damping. In all cases examined numerically, the $\mathcal{S}_0$-bifurcation at $\varepsilon = \lambda_1^{(0)}$ precedes the $\mathcal{S}_1$-bifurcation at $\varepsilon = \varepsilon^\mu$, so the radial sector always governs the onset of instability.
    \item \emph{Dissipation-independent stable windows ($\varepsilon_{\text{spike}}$):} Narrow, discrete stable regions enclosed below by $\varepsilon_{\text{spike}}$ emerge within the parameter space, where the system remains stable independently of the dissipation parameter.
    \item \emph{Large $\kappa$:} There exists a maximum value of $\kappa$ beyond which the stability zone disappears entirely. In dimension $N = 1$, this occurs at $\kappa \approx 4.0$; for higher dimensions, the threshold decreases.
\end{itemize}

Compared with the one-dimensional studies of \citet{barashenkov:1991,carreno:2024}, the arbitrary-dimensional analysis reveals a genuinely new radial instability mechanism for $\kappa < 2/N$, a dimension-dependent maximal stabilisable exponent $\kappa_{\max}^N$, and the existence of narrow, dissipation-independent stable windows bounded below by $\varepsilon_{\text{spike}}$. For the cubic problem, our numerical results are consistent with the two-dimensional oscillons of \citet{barashenkov:2002}, but they go further by identifying the responsible angular sector analytically and by showing that the three-dimensional cubic soliton admits no stability window. 

These results fully map the spectral thresholds governing soliton stability across arbitrary dimensions. Extending this framework to analytically characterise the formation of complex quadruplets stands as a key open challenge for future research.
	\paragraph{Acknowledgements}
	I am grateful for the assistance of Professor Niurka R. Quintero and Professor Renato Alvarez-Nodarse in the development of this work.

	\paragraph{Funding statement}
	The author was supported by FPU24/00951 (Ministerio de Ciencia, Innovaci\'on y Universidades, Spain), by FQM-415 (Plan Propio, Universidad de Sevilla, Spain), and by the IMUS Mar\'ia de Maeztu grant CEX2024-001517-M (Apoyo a Unidades de Excelencia Mar\'ia de Maeztu), funded by MICIU/AEI/10.13039/501100011033.

	\paragraph{Conflict of interest statement} The author declares that he has no competing financial interests or personal
	relationships that could have influenced the work reported in this paper.

	\paragraph{Data availability statement} The code used to generate the numerical results in this paper is available upon request.
	\printbibliography

\appendix

\section{Complex quadruplets in the subcritical regime}\label{app:subcritical_quadruplets}

We prove here the statement used in the numerical discussion: in the subcritical regime, every complex quadruplet automatically satisfies the necessary stability condition~\eqref{neccesary_condition_stability}.

\begin{proposition}[Subcritical complex quadruplets]
	Assume $0 < \kappa < 2/N$ and let $0 < \varepsilon < \lambda_1^{(0)}$. If the eigenvalue problem~\eqref{eq:9kappa}--\eqref{eq:9kappaa} restricted to $\mathcal{S}_0$ has a complex eigenvalue
	\[
		\Lambda = \Lambda_R + i\Lambda_I, \qquad \Lambda_R\Lambda_I \neq 0,
	\]
	then
	\[
		|\Lambda_I| > |\Lambda_R|.
	\]
\end{proposition}

\begin{proof}
	Since $L_0^0$ is invertible, the radial eigenvalue problem can be rewritten as
	\begin{equation}\label{eq:appendix_quad_pencil}
		L_1 f = -\Lambda^2 L_0^{-1} f,
	\end{equation}
	with $f \in \mathcal{S}_0 \setminus \{0\}$. Decompose
	\begin{equation}
		f = \alpha\phi + f^{\perp}, \qquad \alpha \in \mathbb{C}, \qquad f^{\perp} \in \mathcal{S}_0 \cap \{\phi\}^{\perp}.
	\end{equation}
	Since $\tilde{L}_0\phi = 0$, we have $L_0\phi = -\varepsilon\phi$, hence
	\begin{equation}
		L_0^{-1}\phi = -\frac{1}{\varepsilon}\phi.
	\end{equation}
	Moreover, because $\phi$ spans an eigenspace of the self-adjoint operator $L_0$, the subspace $\{\phi\}^{\perp}$ is invariant under $L_0^{-1}$. As $L_0^0$ has exactly one negative eigenvalue, $L_0^{-1}$ is positive on $\mathcal{S}_0 \cap \{\phi\}^{\perp}$.

	Projecting~\eqref{eq:appendix_quad_pencil} onto $\phi$ gives
	\begin{equation}\label{eq:appendix_quad_phi}
		\alpha\langle \phi, L_1\phi \rangle + \langle \phi, L_1 f^{\perp} \rangle = \frac{\Lambda^2}{\varepsilon}\,\alpha\,\|\phi\|^2.
	\end{equation}
	Projecting onto $f^{\perp}$ yields
	\begin{equation}\label{eq:appendix_quad_perp}
		\alpha\langle f^{\perp}, L_1\phi \rangle + \langle f^{\perp}, L_1 f^{\perp} \rangle = -\Lambda^2\,\langle f^{\perp}, L_0^{-1} f^{\perp} \rangle.
	\end{equation}
	Using self-adjointness of $L_1$, we have
	\begin{equation}
		\langle f^{\perp}, L_1\phi \rangle = \langle \phi, L_1 f^{\perp} \rangle^\star.
	\end{equation}
	Taking the complex conjugate of~\eqref{eq:appendix_quad_phi}, multiplying by $\alpha$, and subtracting~\eqref{eq:appendix_quad_perp}, we eliminate the mixed term and obtain
	\begin{equation}\label{eq:appendix_quad_combined}
		|\alpha|^2\langle \phi, L_1\phi \rangle - \langle f^{\perp}, L_1 f^{\perp} \rangle = \frac{{\Lambda^2}^\star}{\varepsilon}|\alpha|^2\|\phi\|^2 + \Lambda^2\langle f^{\perp}, L_0^{-1} f^{\perp} \rangle.
	\end{equation}

	The left-hand side of~\eqref{eq:appendix_quad_combined} is real. Taking imaginary parts gives
	\begin{equation}
		\Im(\Lambda^2)\left(\langle f^{\perp}, L_0^{-1} f^{\perp} \rangle - \frac{|\alpha|^2}{\varepsilon}\|\phi\|^2\right)=0.
	\end{equation}
	For a genuine complex quadruplet, $\Lambda_R\Lambda_I \neq 0$, so
	\begin{equation}
		\Im(\Lambda^2) = 2\Lambda_R\Lambda_I \neq 0.
	\end{equation}
	Hence every complex quadruplet must satisfy the constraint
	\begin{equation}\label{eq:appendix_quad_constraint}
		\langle f^{\perp}, L_0^{-1} f^{\perp} \rangle = \frac{|\alpha|^2}{\varepsilon}\|\phi\|^2.
	\end{equation}
	Since $L_0^{-1} > 0$ on $\mathcal{S}_0 \cap \{\phi\}^{\perp}$, this also shows that any complex quadruplet must have $\alpha \neq 0$ and $f^{\perp} \neq 0$.

	Using the above condition in \eqref{eq:appendix_quad_combined}, we obtain
	\begin{equation}\label{eq:appendix_quad_ReLambda}
		\frac{2\Re(\Lambda^2)}{\varepsilon}|\alpha|^2\|\phi\|^2 = |\alpha|^2\langle \phi, L_1\phi \rangle - \langle f^{\perp}, L_1 f^{\perp} \rangle.
	\end{equation}

	We now determine the sign of the right-hand side. Since
	\begin{equation}
		L_1\phi = -2\kappa\phi^{2\kappa+1},
	\end{equation}
	we have
	\begin{equation}
		\langle \phi, L_1\phi \rangle = -2\kappa\int_{\mathbb{R}^N}\phi^{2\kappa+2}\,d\bx < 0.
	\end{equation}
	Also, in the subcritical regime $\kappa < 2/N$, the orthogonality result for $L_1$ on $\{\phi\}^{\perp}$ (see Lemma~\ref{lem:L1_spectral}(iv), following \citet{weinstein:1985}) yields
	\begin{equation}
		\langle f^{\perp}, L_1 f^{\perp} \rangle > 0
	\end{equation}
	for every nonzero $f^{\perp} \in \mathcal{S}_0 \cap \{\phi\}^{\perp}$. Therefore the right-hand side of~\eqref{eq:appendix_quad_ReLambda} is strictly negative, while its prefactor is strictly positive. We conclude that
	\begin{equation}
		\Re(\Lambda^2) < 0,
	\end{equation}
	this is equivalent to
	\begin{equation}
		\Lambda_R^2 - \Lambda_I^2 < 0,
	\end{equation}
	hence $|\Lambda_I| > |\Lambda_R|$.
\end{proof}

\section{Eigenvalues and eigenfunctions of \texorpdfstring{$L_0$ and $L_1$}{L0 and L1} in dimension one}\label{app:eigenvalues_dim1}

In dimension $N = 1$, the operators $L_0$ and $L_1$ admit a complete analytic solution. We describe the method for $\tilde{L}_0 g = \lambda g$; the analysis for $L_1$ is analogous; see, for example, \citep{raban:2022}.

Recall that $\tilde{L}_0 = -\partial_x^2 + 1 - (\kappa+1)\operatorname{sech}^2(\kappa x)$. The change of variables $y = \tanh(\kappa x)$ transforms the eigenvalue equation into an associated Legendre equation:
\begin{equation}
	-\frac{\partial}{\partial y}\!\left((1-y^2)\frac{\partial g}{\partial y}\right) + \frac{1-\lambda}{\kappa^2}\frac{1}{1-y^2}g - \frac{\kappa+1}{\kappa^2}g = 0.
\end{equation}
Seeking solutions $g(y) = (1-y^2)^\alpha h(y)$ with $h$ bounded and $h(1) \neq 0$, the behaviour at $y = 1$ requires $4\alpha^2 = (1-\lambda)/\kappa^2$. Expanding $h(y) = \sum_{j=0}^\infty A_j y^j$ yields the recurrence
\begin{equation}
	(n+2)(n+1)A_{n+2} = \left(n(n-1) + 2n(2\alpha+1) + 4\alpha^2 + 2\alpha - \frac{\kappa+1}{\kappa^2}\right)A_n.
\end{equation}
By the Raabe criterion the series diverges unless it terminates. If the series stops at degree $n$, then the admissible branch is determined by
\[
	2\alpha + n = \frac{1}{\kappa}, \qquad \alpha = \frac{1/\kappa - n}{2}.
\]
The other algebraic branch, $2\alpha + n = -1 - 1/\kappa$, is discarded because it yields $\alpha < 0$ and therefore non-$L^2(\mathbb{R})$ solutions growing at infinity.
Therefore the discrete spectrum occurs only for integers $n$ such that $0 \le n < 1/\kappa$, and the corresponding eigenvalues are
\begin{equation}\label{eq:eigenvalues_tildeL0}
	\lambda_n = \kappa n(2 - \kappa n), \qquad n \in \mathbb{N}_0,\quad n < \frac{1}{\kappa}.
\end{equation}
with eigenfunctions
\begin{equation}
	g_n(x) = \operatorname{sech}^{1/\kappa-n}(\kappa x)\sum_{m=0}^n A_{m,\lambda_n}\tanh^m(\kappa x).
\end{equation}
In particular: $\lambda_0 = 0$ with $g_0 = \phi$; $\lambda_1 = \kappa(2-\kappa)$, which belongs to the discrete spectrum only for $\kappa < 1$ and has odd parity; and $\lambda_2 = 4\kappa(1-\kappa)$, which belongs to the discrete spectrum only for $\kappa < 1/2$ and has even parity.

For $L_1$, the eigenvalues are $\mu_n = \kappa(n-1)(2-\kappa(n-1))$, the same as $\tilde{L}_0$ but shifted in~$n$, with admissible indices $n \in \mathbb{N}_0$ such that $n < 1 + 1/\kappa$. In particular, $\mu_0 = -\kappa(2+\kappa)$ with eigenfunction $f_0 = \phi^{\kappa+1}$, and $\mu_1 = 0$ with eigenfunction $f_1 = \phi_x$.

The operator $L_0 = \tilde{L}_0 - \varepsilon$ has eigenvalues $\lambda_n - \varepsilon$ with the same eigenfunctions.

\end{document}